\documentclass[sn-mathphys,Numbered]{sn-jnl}

\usepackage{graphicx}%
\usepackage{multirow}%
\usepackage{amsmath,amssymb,amsfonts}%
\usepackage{amsthm}%
\usepackage{mathrsfs}%
\usepackage[title]{appendix}%
\usepackage{xcolor}%
\usepackage{textcomp}%
\usepackage{manyfoot}%
\usepackage{booktabs}%
\usepackage{algorithm}%
\usepackage{algorithmicx}%
\usepackage{algpseudocode}%
\usepackage{listings}%

\theoremstyle{thmstyleone}%
\theoremstyle{thmstyletwo}%

\theoremstyle{thmstylethree}%

\begin{document}

\title[High critical current densities of bcc HEA superconductors: recent research progress]{High critical current densities of body-centered cubic high-entropy alloy superconductors: recent research progress}


\author*[1]{\fnm{Jiro} \sur{Kitagawa}}\email{j-kitagawa@fit.ac.jp}

\author[2]{\fnm{Yoshikazu} \sur{Mizuguchi}}

\author[3]{\fnm{Terukazu} \sur{Nishizaki}}


\affil[1]{\orgdiv{Department of Electrical Engineering, Faculty of Engineering}, \orgname{Fukuoka Institute of Technology}, \orgaddress{\street{3-30-1 Wajiro-higashi, Higashi-ku}, \city{Fukuoka}, \postcode{811-0295}, \country{Japan}}}

\affil[2]{\orgdiv{Department of Physics}, \orgname{Tokyo Metropolitan University}, \orgaddress{\street{1-1, Minami-osawa}, \city{Hachioji}, \postcode{192-0397}, \country{Japan}}}

\affil[3]{\orgdiv{Department of Electrical Engineering, Faculty of Science and Engineering}, \orgname{Kyushu Sangyo University}, \orgaddress{\street{2-3-1 Matsukadai, Higashi-ku}, \city{Fukuoka}, \postcode{813-8503}, \country{Japan}}}


\abstract{High-entropy alloy (HEA) superconductors have garnered significant attention due to their unique characteristics, such as robust superconductivity under extremely high pressure and irradiation, the cocktail effect, and the enhancement of the upper critical field. A high critical current density is another noteworthy feature observed in HEAs. Several body-centered cubic (bcc) HEAs have exhibited critical current densities comparable to those of Nb-Ti superconducting alloys. Such HEAs hold potential for applications as multifunctional superconducting wires, a capability rarely achieved in conventional alloys. In this context, we review recent advancements in research on critical current densities in bcc HEA superconductors, including Ta$_{1/6}$Nb$_{2/6}$Hf$_{1/6}$Zr$_{1/6}$Ti$_{1/6}$, (TaNb)$_{0.7}$(HfZrTi)$_{0.5}$, NbScTiZr, and others. Comparative analyses among these HEAs reveal that both eutectic microstructures, which accompany lattice strain, and nanosized precipitates play pivotal roles in achieving elevated critical current densities across wide magnetic field ranges. Furthermore, we propose several future directions for research. These include elucidating the origin of lattice strain, exploring more fine eutectic microstructures, artificially introducing nanoscale pinning sites, improving the superconducting critical temperature, and investigating the mechanical properties of these materials.}

\maketitle

\section{Introduction}\label{sec1}
High-entropy alloys (HEAs) have garnered significant attention in materials science due to their diverse and versatile functionalities, including exceptional mechanical strength, corrosion resistance, energy storage capabilities, magnetic refrigeration, soft ferromagnetism, catalytic potential, and thermoelectricity\cite{Gao:book,Biswas:book,Li:PMS2021,Fu:JMST2021,Marques:EES2021,Law:JMR2023,Nakamura:AIPAd2023,Chaudhary:MT2021,Kitagawa:APLMater2022,Kitagawa:JMMM2022,Li:ME2021,Jiang:Science2021}. 
The defining feature of HEAs lies in the incorporation of multiple principal elements with high mixing entropy values. 
According to thermodynamic theory, higher mixing entropy stabilizes solid solution phases, particularly at elevated temperatures. 
Many researchers adopt a threshold mixing entropy value of 1.0 $R$ ($R$: gas constant) for HEAs, typically achieved by including at least four constituent elements\cite{Yan:MMTA2021}.
The HEA concept has been extensively explored across various materials, with high-entropy materials broadly classified into two categories: high-entropy alloys and high-entropy compounds. 
Unlike conventional alloys, which are predominantly composed of one or two principal elements, HEAs exhibit distinctive "four core effects": the high-entropy effect, severe lattice distortion, sluggish diffusion, and the cocktail effect\cite{Gao:book,Biswas:book}. 
Another remarkable characteristic of HEAs is their multifunctionality\cite{Han:NRM2024}. 
Multiple functionalities can coexist in HEAs, such as the combination of high strength and ductility or soft magnetism and corrosion resistance\cite{Lu:SM2020,Duan:SCM2023}, which is rarely achieved in conventional alloys.

Superconductivity has emerged as one of the intriguing functionalities explored in HEAs since the discovery of the body-centered cubic (bcc) Ta$_{34}$Nb$_{33}$Hf$_{8}$Zr$_{14}$Ti$_{11}$ superconductor in 2014\cite{Kozelj:PRL2014,Sun:PRM2019,Kitagawa:Metals2020,Kitagawa:book,Zeng:NPGAM2024}. 
Presently, HEA superconductivity is studied across diverse structural types, including bcc\cite{Rohr:PRM2018,Marik:JALCOM2018,Ishizu:RINP2019,Harayama:JSNM2021,Sarkar:IM2022,Motla:PRB2022,Kitagawa:RHP2022,Kitagawa:JALCOM2022,Hattori:JAMS2023,Li:JPCC2023,Zeng:SCPMA2023}, hexagonal close-packed (hcp)\cite{Lee:PhysicaC2019,Marik:PRM2019,Browne:JSSC2023}, face-centered cubic (fcc)\cite{Zhu:JALCOM2022,Strong:EM2024}, CsCl-type\cite{Stolze:ChemMater2018}, A15\cite{Yamashita:JALCOM2021}, NaCl-type\cite{Mizuguchi:JPSJ2019}, $\alpha$ (or $\beta$)-Mn-type\cite{Stolze:JMCC2018,Xiao:SM2023}, $\sigma$-phase type\cite{Liu:ACS2020}, CuAl$_{2}$-type\cite{Kasen:SST2021}, NiAs-type\cite{Hirai:CM2024}, BiS$_{2}$-based\cite{Sogabe:SSC2019}, and YBCO-based\cite{Shukunami:PhysicaC2020} structures.
The bcc, hcp, and fcc HEAs are classified as alloy-type, while the remaining structures fall under high-entropy compounds.
Several key findings have been reported in HEA superconductors, including the robustness of superconductivity against high pressures and irradiation, the cocktail effect, and the enhancement of the upper critical field induced by the high-entropy effect\cite{Guo:PNAC2017,Jung:NC2022,Sogabe:SSC2019,Hirai:CM2024,Kasem:SciRep2022,Jangid:APL2024,Nakahira:JMS2022}.
For instance, the bcc (TaNb)$_{0.67}$(HfZrTi)$_{0.33}$ alloy demonstrates the robustness of superconductivity under high pressures\cite{Guo:PNAC2017}, with a superconducting critical temperature ($T_\mathrm{c}$=8 K) remaining constant even under extreme pressures of up to 180 GPa.
The cocktail effect, one of the four core effects, refers to a synergistic phenomenon where physical properties surpass the average values predicted by the mixture rule. 
This effect has been observed in BiS$_{2}$-based superconductors, where increasing mixing entropy enhances the diamagnetic signal\cite{Sogabe:SSC2019}. 
More recently, an enhancement of the zero-temperature upper critical field, $\mu_{0}H_\mathrm{c2}$(0), with increasing mixing entropy has been reported in NiAs-type (RuRhPdIr)$_{1-x}$Pt$_{x}$Sb\cite{Hirai:CM2024}. 
As the Pt content decreases from $x$=1 to $x$=0.2, $\mu_{0}H_\mathrm{c2}$(0) increases significantly from 0.07 T to 1.33 T.

\begin{figure}
\begin{center}
\includegraphics[width=0.6\linewidth]{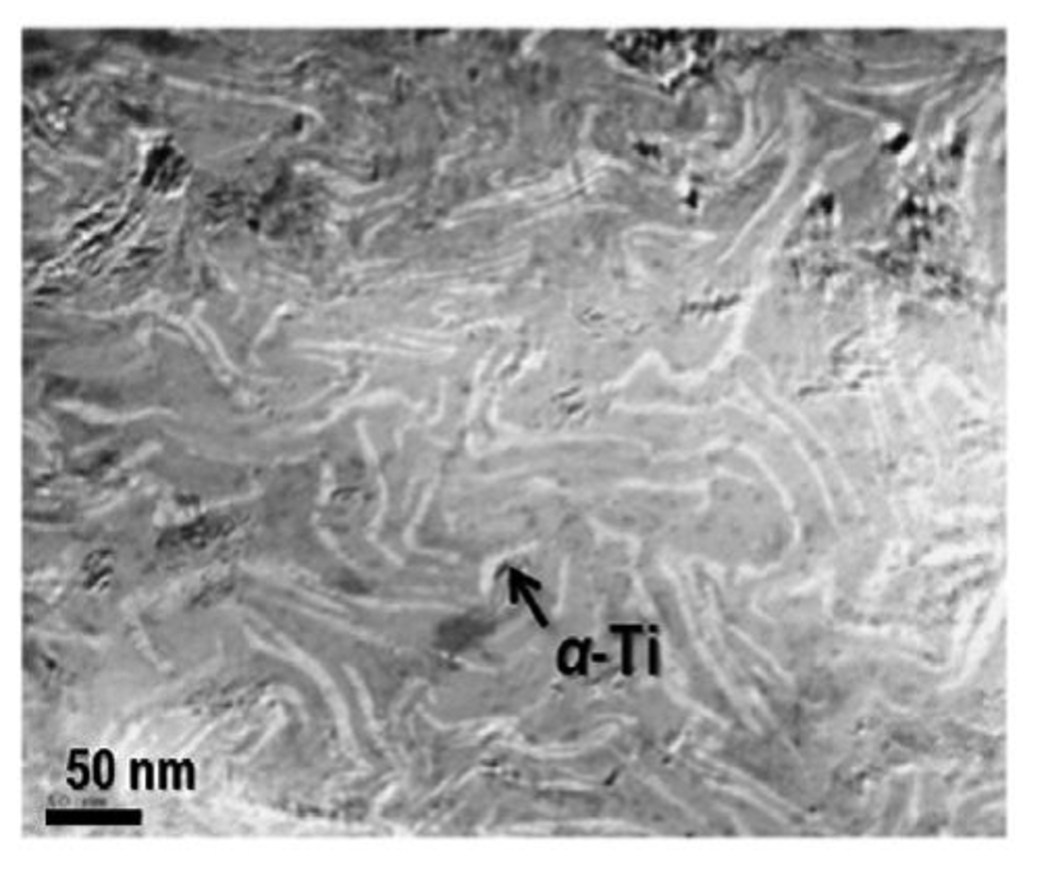}
\caption{\label{fig1} Microstructure of Nb–47 wt.\%Ti alloy. Reproduced with permission from \cite{Li:Physc2008}.}
\end{center}
\end{figure}

The exploration of practical applications for HEA superconductors is critically important for advancing superconducting technology. 
Notable properties such as high critical current density, exceptional hardness, and the shape memory effect are impressive\cite{Jung:NC2022,Kitagawa:JALCOM2022,Egilmez:PRM2021}.
The latter two features are especially promising for developing multifunctional superconducting devices\cite{Kitagawa:JALCOM2022,Egilmez:PRM2021}.
Several body-centered cubic (bcc) HEAs have demonstrated critical current densities ($J_\mathrm{c}$) comparable to those of commercial Nb-Ti superconducting wires\cite{Jung:NC2022,Gao:APL2022,Seki:JSNM2023}. 
In conventional bcc superconducting alloys, Nb–47 wt.\%Ti is widely utilized for superconducting wires\cite{Froes:book2019,Banno:Super2023}. 
The fabrication process typically involves heat treatments ranging from 380 to 420 $^{\circ}$C for 40 to 80 hours\cite{Froes:book2019,Banno:Super2023}. 
Following the final heat treatment, a strain of 330–450 \% is applied\cite{Froes:book2019,Banno:Super2023}. 
This intricate fabrication procedure induces nanoscale $\alpha$-Ti precipitates that serve as effective pinning sites. 
An example of this microstructure is shown in Fig.\hspace{1mm}\ref{fig1}\cite{Li:Physc2008}, where $\alpha$-Ti forms ribbon-like features approximately 5 nm thick, with 10–30 nm spacings. 
This nanoscale microstructure underpins the high $J_\mathrm{c}$ values, exceeding 10$^{5}$ A/cm$^{2}$, which meet the practical requirements\cite{Larbalestier:Nature2001,Komori:APL2002,Jung:NC2022}, even under elevated magnetic fields.
The relationships between $J_\mathrm{c}$ and microstructural features have also been examined in recent studies on bcc HEA superconductors\cite{Gao:APL2022,Seki:JSNM2023,Kim:AM2022}, a central focus of this review. 
Additionally, as previously noted, the unique multifunctionality of HEAs is rarely observed in conventional alloys such as Nb-Ti. 
The bcc HEA superconductors exhibiting high $J_\mathrm{c}$ values discussed in this review show considerable promise for practical applications in extreme environments, such as aerospace and nuclear fusion reactors, owing to the inherent high irradiation resistance of HEAs\cite{Shi:Tungsten2021}. 
Consequently, investigating $J_\mathrm{c}$ of bcc HEA superconductors remains a vital area of research.

In this review, we focus on bcc HEA superconductors and summarize recent progress in high-$J_\mathrm{c}$ bcc HEAs, including Ta$_{1/6}$Nb$_{2/6}$Hf$_{1/6}$Zr$_{1/6}$Ti$_{1/6}$, (TaNb)$_{0.7}$(HfZrTi)$_{0.5}$, NbScTiZr, and others. 
For each system, we discuss the fundamental superconducting properties and the magnetic field performance of $J_\mathrm{c}$. 
A comparative analysis of $J_\mathrm{c}$ among the representative bcc HEAs surveyed in this review is also presented. 
Finally, we address future perspectives on the practical application of these materials as superconducting wires.

\section{Bean model and pinning mechanisms}\label{sec2}
In this section, we provide a concise overview of the Bean model for extracting $J_\mathrm{c}$ and summarize the flux pinning mechanisms. 
Numerous studies cited in this review frequently reference the Bean model and the analysis of pinning mechanisms.

The Bean model predicts the response of a superconductor with $J_\mathrm{c}$ to variations in the external magnetic field under isothermal conditions and explains the magnetic hysteresis loop\cite{Bean:RMP1964}. 
This model assumes that flux motion within a sample induces a local current density of $J_\mathrm{c}$.
When the external magnetic field is parallel to the sample surface, the magnetic field within the sample decreases linearly with distance from the surface, as dictated by Amp\`{e}re's law. 
Consequently, $J_\mathrm{c}$ is expressed as follows:
\begin{equation}
J_\mathrm{c}=\frac{20\Delta M}{a\left(1-a/3b\right)}
\label{eq:bean}
\end{equation}
where $\Delta M$ represents the vertical width of the magnetic hysteresis loop, and $a$ and $b$ ($a$$<$$b$) denote the sample dimensions perpendicular to the external field. 
Thus, a larger hysteresis loop area corresponds to a higher $J_\mathrm{c}$.

To elucidate the pinning mechanism, the first step involves calculating the magnetic field dependence of the flux pinning force density, defined as $F_\mathrm{p}=\left| \mu_{0}\vec{H}\times\vec{J_\mathrm{c}}\right|$, where $\mu_{0}$ is the vacuum permeability, and $H$ is the external magnetic field. 
Next, the normalized flux pinning force density, $f_\mathrm{p}$ ($=F_\mathrm{p}/F_\mathrm{p,max}$), is computed as a function of the reduced field $h$ ($=H/H^\mathrm{*}$) at various temperatures. 
Here, $F_\mathrm{p,max}$ denotes the maximum value of $F_\mathrm{p}$, and $H^\mathrm{*}$ represents the irreversible field. 
The functional form of $f_\mathrm{p}$ depends on the specific pinning model. 
The Dew-Hughes model\cite{Dew-Hughes:PhyMag1974} provides a general expression for $f_\mathrm{p}$ as $\propto h^{p}(1-h)^{q}$, where $p$ and $q$ are determined by the nature of the flux pinning in the superconductor. 
Representative pinning mechanisms discussed in this review are as follows:

\begin{itemize}
      \item normal surface pinning: $f_\mathrm{p}\propto h^{0.5}(1-h)^{2}$
      \item normal point pinning: $f_\mathrm{p}\propto h(1-h)^{2}$
      \item $\Delta\kappa$ pinning: $f_\mathrm{p}\propto h^{2}(1-h)$
\end{itemize}

In the normal surface pinning and normal point pinning models, grain boundaries and precipitates serve as pinning centers, respectively. 
If spatial variations in the Ginzburg-Landau parameter $\kappa$ arise due to compositional fluctuations in the superconductor, these lead to flux pinning described by the $\Delta\kappa$ pinning model.
The functional form of $f_\mathrm{p}$ is often scaled with $h'=H/H_\mathrm{p}$ instead of $h$. 
Here, $H_\mathrm{p}$ represents the magnetic field corresponding to $F_\mathrm{p,max}$. 
In this context, the scaling functions are typically modified\cite{Higuchi:PRB1999,Shigeta:PhysC2003,Cai:APL2013}  as follows: $f_\mathrm{p}=25/16h'^{0.5}(1-h'/5)^{2}$ for normal surface pinning, $f_\mathrm{p}=9/4h'(1-h'/3)^{2}$ for normal point pinning, and  $f_\mathrm{p}=3h'^{2}(1-2h'/3)$ for $\Delta\kappa$ pinning.
These formulations are frequently applied to HEA superconductors. 
To model the magnetic field performance of $J_\mathrm{c}$ in two-gap superconductors, the double exponential model is employed\cite{Jung:JAP2013}. 
In this model, $J_\mathrm{c}(h)$ is described by $J_{1}\exp(-A_{1}h)+J_{2}\exp(-A_{2}h)$ where $A_{1}$ and $A_{2}$ are proportionality constants, and $J_{1}$ and $J_{2}$ are the critical current densities corresponding to the large and small gaps, respectively.

\section{Ta$_{1/6}$Nb$_{2/6}$Hf$_{1/6}$Zr$_{1/6}$Ti$_{1/6}$}\label{sec3}
The HEA, composed of refractory elements, is commonly known as the Senkov alloy.
The Ta-Nb-Hf-Zr-Ti Senkov alloy, exhibiting a single bcc structure, has garnered significant attention due to its exceptional room-temperature tensile ductility\cite{Senkov:JMS2012}.
This property may be advantageous for the fabrication of superconducting wires.
Ta-Nb-Hf-Zr-Ti is a well-studied superconducting HEA, and the cocktail effect of $T_\mathrm{c}$ and the robust superconductivity against high pressure or irradiation distinguish Ta-Nb-Hf-Zr-Ti from conventional alloys.

The $J_\mathrm{c}$ characteristics of Ta$_{1/6}$Nb$_{2/6}$Hf$_{1/6}$Zr$_{1/6}$Ti$_{1/6}$ have been extensively studied by the research teams at Kyung Hee University and Sungkyunkwan University. 
Readers are encouraged to refer to the comprehensive review\cite{Hidayati:CAP2024}, which partially discusses the $J_\mathrm{c}$ results of Ta$_{1/6}$Nb$_{2/6}$Hf$_{1/6}$Zr$_{1/6}$Ti$_{1/6}$.

\subsection{As-cast bulk sample prepared by arc-melting technique}
The first report on the $J_\mathrm{c}$ of Ta$_{1/6}$Nb$_{2/6}$Hf$_{1/6}$Zr$_{1/6}$Ti$_{1/6}$ was published in 2020\cite{Kim:AM2020}. 
This HEA with a single-phase structure was synthesized using a conventional arc-melting technique, and the superconducting properties of the as-cast state were evaluated. 
The fundamental superconducting properties were characterized through measurements of magnetization, electrical resistivity, and specific heat. 
The HEA undergoes a superconducting transition at 7.85 K. 
Analysis of the temperature dependence of the upper critical field suggests a $\mu_{0}H_\mathrm{c2}$(0) of 12.05 T, from which the zero-temperature Ginzburg–Landau coherence length, $\xi_\mathrm{GL}$(0), is calculated to be 5.23 nm.

\begin{figure}
\begin{center}
\includegraphics[width=0.6\linewidth]{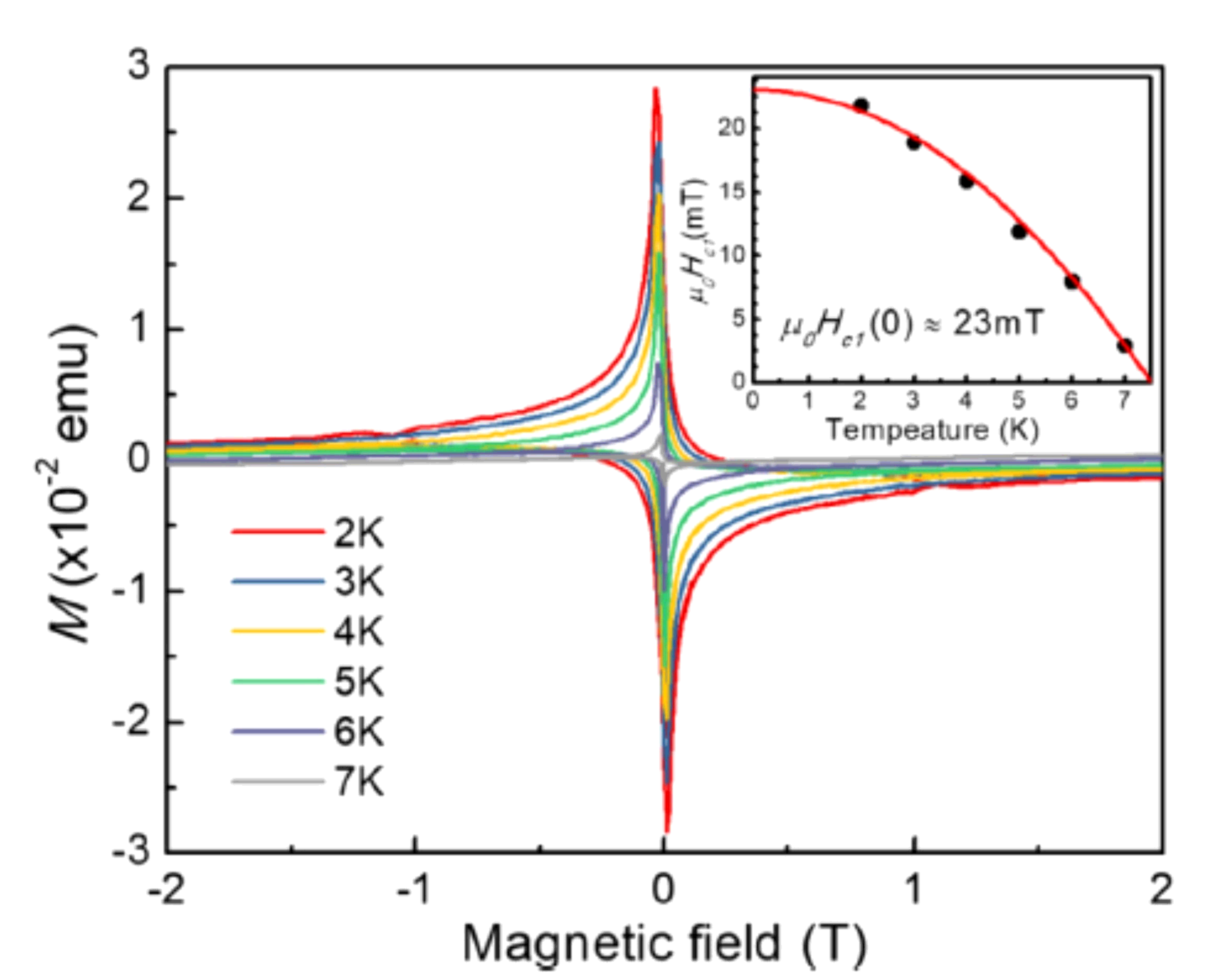}
\caption{\label{fig2} Isothermal magnetization curves of as-cast Ta$_{1/6}$Nb$_{2/6}$Hf$_{1/6}$Zr$_{1/6}$Ti$_{1/6}$. Reproduced with permission from \cite{Kim:AM2020}.}
\end{center}
\end{figure}

\begin{figure}
\begin{center}
\includegraphics[width=1\linewidth]{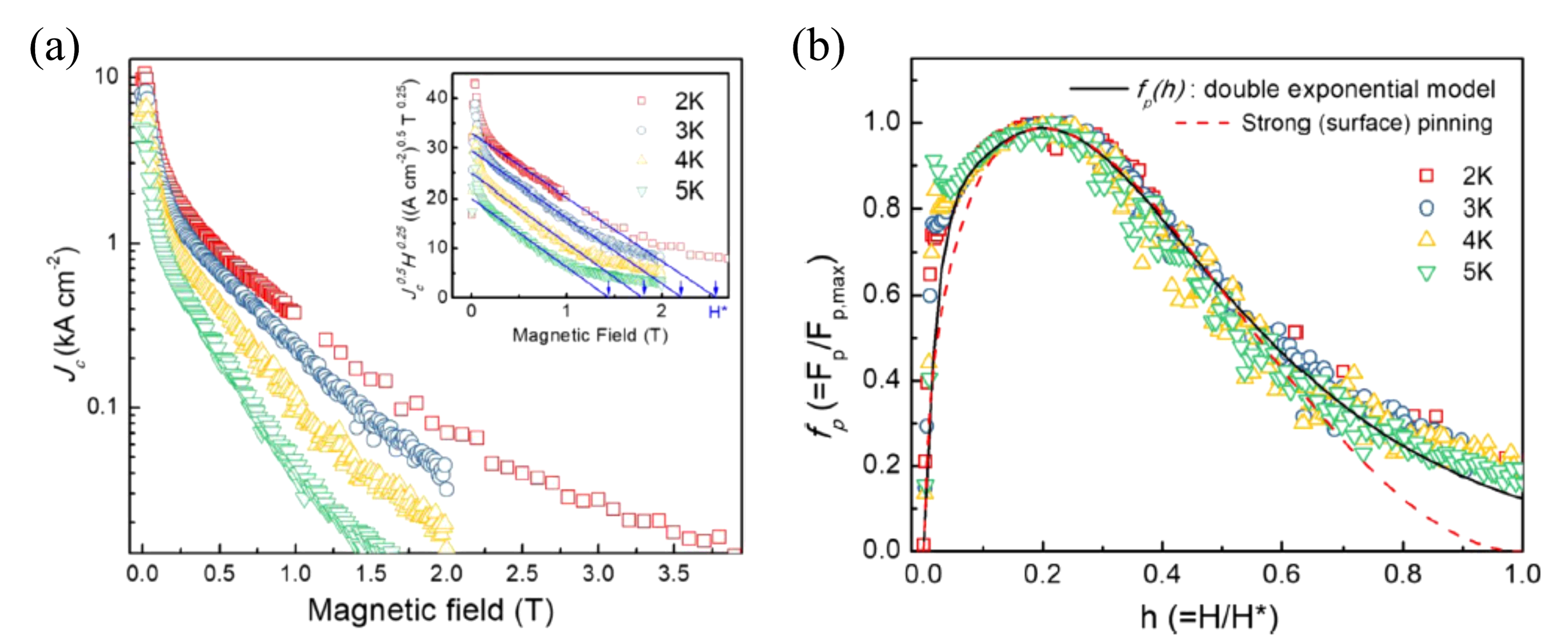}
\caption{\label{fig3} (a) Field-dependent $J_\mathrm{c}$ and (b) normalized flux pinning force density $f_\mathrm{p}$ as a function of the reduced field $h$ for as-cast Ta$_{1/6}$Nb$_{2/6}$Hf$_{1/6}$Zr$_{1/6}$Ti$_{1/6}$. Reproduced with permission from \cite{Kim:AM2020}.}
\end{center}
\end{figure}

Figure\hspace{1mm}\ref{fig2} displays the magnetic hysteresis loops measured at temperatures ranging from 2 to 7 K. 
Figure\hspace{1mm}\ref{fig3}(a) presents the magnetic field dependence of $J_\mathrm{c}$ measured between 2 and 5 K. 
All $J_\mathrm{c}$ data demonstrate the conventional trend of decreasing $J_\mathrm{c}$ with increasing magnetic field. 
$J_\mathrm{c}$ systematically increases across the entire magnetic field range as the temperature decreases. 
The self-field $J_\mathrm{c}$ of 10,655 A/cm$^{2}$ at 2 K represents the highest value observed; however, it remains below the practical threshold of 10$^{5}$ A/cm$^{2}$.
The critical exponent $\alpha$ in the magnetic field dependence of $J_\mathrm{c}$ expressed as $\propto H^{-\alpha}$ ($H$: external field) is 0.85 at 2 K, indicating that collective pinning occurs within the sample.
Figure\hspace{1mm}\ref{fig3}(b) plots the normalized flux pinning force density, $f_\mathrm{p}$, as a function of the reduced field, $h$, at various temperatures. 
The magnetic field dependence of $f_\mathrm{p}$ is well described by the surface pinning model or the double exponential model. 
The former suggests that grain boundaries in the polycrystalline sample act as flux pinning centers. 
However, the as-cast Ta$_{1/6}$Nb$_{2/6}$Hf$_{1/6}$Zr$_{1/6}$Ti$_{1/6}$ with its single-phase structure appears inadequate for practical applications as a superconducting wire due to the absence of strong flux pinning centers.

\subsection{Bulk sample prepared by spark plasma sintering}\label{subsecSPS}
Subsequently, the Korean research team employed a combination of planetary ball milling and spark plasma sintering (SPS) as the sample preparation method\cite{Kim:AM2022}. 
Powders of the constituent elements, in their stoichiometric ratios, were loaded into a stainless-steel jar under an argon atmosphere and subjected to ball milling. 
SPS was then performed on the compacted powders.
$T_\mathrm{c}$ of the SPS sample was determined to be 7.83 K, which is nearly identical to that of the as-cast bulk sample. 
$\mu_{0}H_\mathrm{c2}$(0) was measured as 10.5 T, which is lower than that of the arc-melted sample (12.05 T). 
The calculated $\xi_\mathrm{GL}$(0) for the SPS sample was 5.66 nm.
The X-ray diffraction pattern of the SPS sample revealed impurity phases of Hf$_{2}$Fe and ZrFe$_{2}$. 
Contamination with Fe, originating from the stainless-steel balls and jar used during ball milling, was found to be unavoidable.

\begin{figure}
\begin{center}
\includegraphics[width=0.6\linewidth]{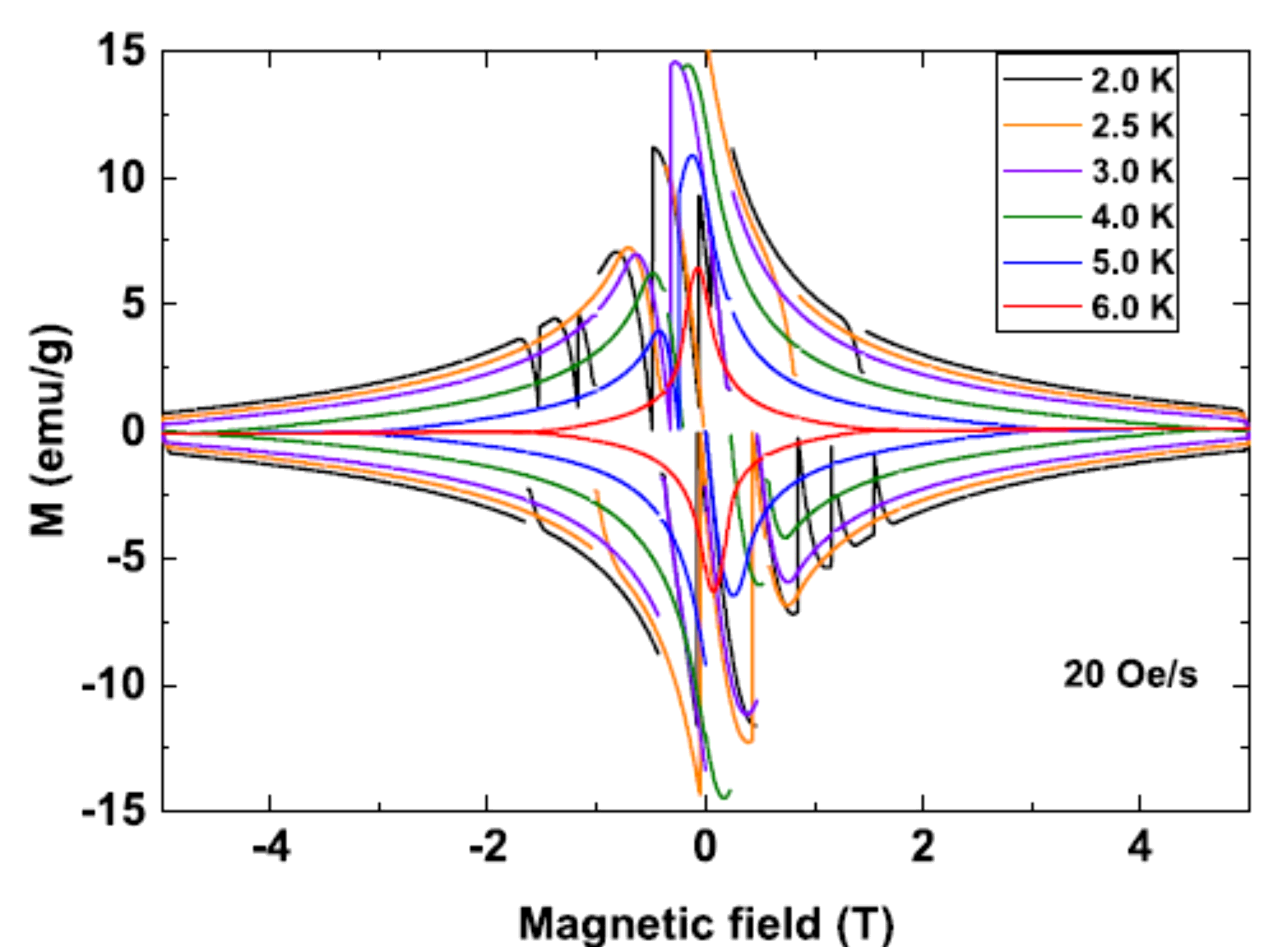}
\caption{\label{fig4} Isothermal magnetic hysteresis loops of Ta$_{1/6}$Nb$_{2/6}$Hf$_{1/6}$Zr$_{1/6}$Ti$_{1/6}$ fabricated via ball-milling and spark plasma sintering (SPS) methods. Reproduced with permission from \cite{Kim:AM2022}.}
\end{center}
\end{figure}

\begin{figure}
\begin{center}
\includegraphics[width=1\linewidth]{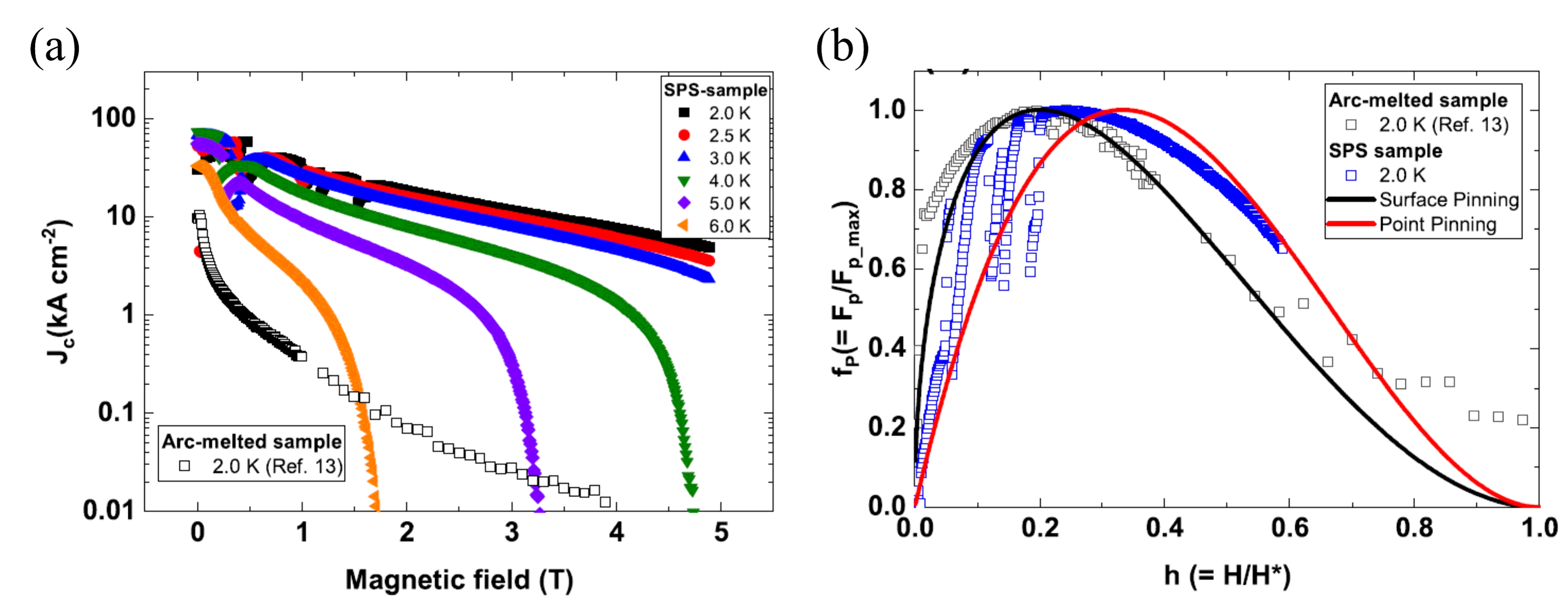}
\caption{\label{fig5} (a) Field-dependent $J_\mathrm{c}$ and (b) normalized flux pinning force density $f_\mathrm{p}$ as a function of the reduced field $h$ for SPS-processed Ta$_{1/6}$Nb$_{2/6}$Hf$_{1/6}$Zr$_{1/6}$Ti$_{1/6}$. Reproduced with permission from \cite{Kim:AM2022}.}
\end{center}
\end{figure}

The isothermal magnetic hysteresis loops measured below $T_\mathrm{c}$ are presented in Fig.\hspace{1mm}\ref{fig4}. 
The loop shape is markedly different from that of the arc-melted sample in the as-cast state. 
The $M$-$H$ ($M$: magnetization, $H$: external field) loop of the SPS sample exhibits magnetization quenching at low magnetic fields, a phenomenon that becomes increasingly pronounced with decreasing temperature. 
This behavior, known as magnetic flux jumps, is characteristic of a superconductor with strong flux pinning and arises from magnetic instability due to disturbances associated with flux motion\cite{Dou:PhysicaC2001}. 
It is noteworthy that magnetic flux jumps are absent in the arc-melted sample, which lacks impurity phases. 
This contrast suggests that microstructural differences between the arc-melted and SPS samples play a crucial role in determining pinning strength.
The $J_\mathrm{c}$ data extracted from magnetic measurements are displayed in Fig.\hspace{1mm}\ref{fig5}(a). 
The $J_\mathrm{c}$ values of the SPS sample are approximately 30,500 A/cm$^{2}$ (at 2 K and 0.01 T) and 73,200 A/cm$^{2}$ (at 4 K and 0.01 T). 
These values represent significant increases of approximately 286 \% (at 2 K and 0.01 T) and 687 \% (at 4 K and 0.01 T) compared to the 10,655 A/cm$^{2}$ (at 2 K and 0.01 T) of the arc-melted sample in the as-cast state. 
The authors attribute this enhancement in $J_\mathrm{c}$ to magnetic flux pinning induced by impurities.
The pinning force density, $F_\mathrm{p}$, of the SPS sample systematically increases as the temperature decreases within the measured magnetic field range.
The maximum $F_\mathrm{p}$ reaches approximately 0.4 GN/m$^{3}$ at 2 K, two orders of magnitude higher than that of the arc-melted sample.
The pinning mechanism was further analyzed by examining $f_\mathrm{p}(h)$, as shown in Fig.\hspace{1mm}\ref{fig5}(b).
The experimental results, including those of the arc-melted sample, were compared with the surface pinning model (black line) and the point pinning model (red line).
While the surface pinning model adequately describes $f_\mathrm{p}(h)$ for the arc-melted sample, the $f_\mathrm{p}(h)$ of the SPS sample likely reflects contributions from both surface and point pinning mechanisms.
The observed enhancement of $J_\mathrm{c}$ in the SPS sample compared to the arc-melted sample is particularly noteworthy, as the preparation method profoundly impacts the $J_\mathrm{c}$ performance. 
In this case, the impurity phases introduced by contamination during the ball milling process appear to be effective magnetic flux pinning sites.

SPS samples exhibit relatively high $J_\mathrm{c}$, primarily due to the introduction of intermetallic impurities. 
The source of contamination is the ball milling process prior to SPS treatment, during which Fe contamination from the stainless-steel balls and jar occurs. 
Enhancing $J_\mathrm{c}$ by combining SPS and ball milling methods may be applicable to other HEAs, provided that the elements used in the balls and jar and the HEA components can form intermetallic compounds.
An alternative approach for introducing intermetallic impurities is the deliberate addition of small amounts of foreign elements during the melting process, which necessitates a thorough understanding of alloy phase formation.

\subsection{Thin film sample}
\begin{figure}
\begin{center}
\includegraphics[width=0.6\linewidth]{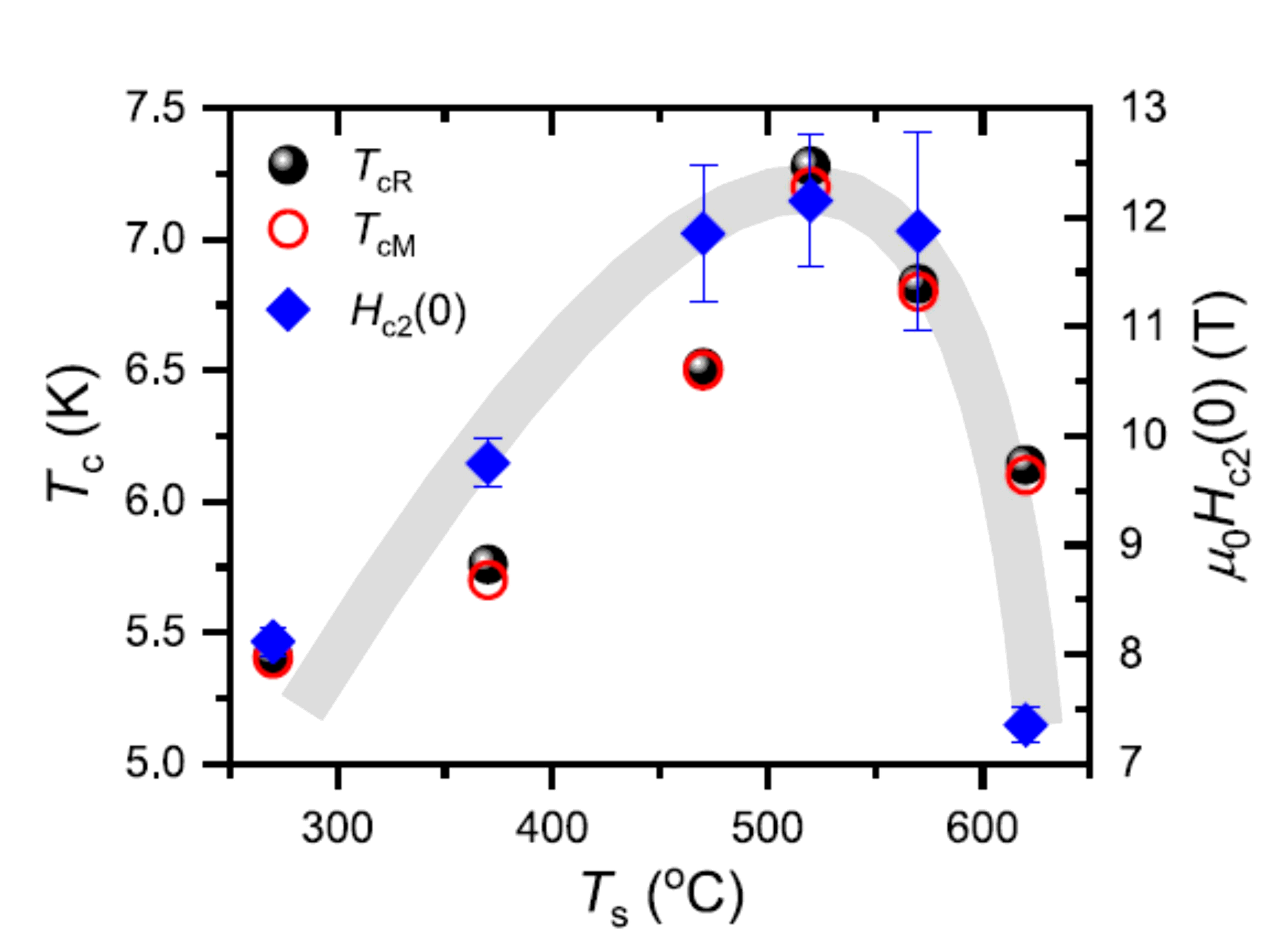}
\caption{\label{fig6} Dependence of $T_\mathrm{c}$ and $\mu_{0}H_\mathrm{c2}$(0) on substrate temperature for thin-film
Ta$_{1/6}$Nb$_{2/6}$Hf$_{1/6}$Zr$_{1/6}$Ti$_{1/6}$. Reproduced from \cite{Jung:NC2022} under Creative Commons Attribution License.}
\end{center}
\end{figure}

The remarkable improvement in $J_\mathrm{c}$ of thin-film Ta$_{1/6}$Nb$_{2/6}$Hf$_{1/6}$Zr$_{1/6}$Ti$_{1/6}$ represents a milestone in the research of HEA superconductors, as reported in Nature Communications in 2022\cite{Jung:NC2022}. 
This study highlights the thin-film Ta$_{1/6}$Nb$_{2/6}$Hf$_{1/6}$Zr$_{1/6}$Ti$_{1/6}$, which exhibits a $J_\mathrm{c}$ significantly exceeding the benchmark performance of Nb-Ti wires used in high-field superconducting magnets. 
The target Ta$_{1/6}$Nb$_{2/6}$Hf$_{1/6}$Zr$_{1/6}$Ti$_{1/6}$ material was synthesized using a ball-milling and hot-press sintering process. 
High-quality thin films were subsequently fabricated on $c$-cut Al$_{2}$O$_{3}$ substrates using a pulsed laser deposition (PLD) technique, with the substrate temperature ($T_\mathrm{s}$) varied between  270 $^{\circ}$C and 620 $^{\circ}$C. 
The thickness of the resulting films ranged from 115 to 700 nm.
The fundamental superconducting properties, $T_\mathrm{c}$ and $\mu_{0}H_\mathrm{c2}$(0), are summarized in Fig.\hspace{1mm}\ref{fig6}. 
The dependence of $T_\mathrm{c}$ on $T_\mathrm{s}$ reveals a dome-like structure, with a maximum $T_\mathrm{c}$ of 7.28 K observed at $T_\mathrm{s}$=520 $^{\circ}$C. 
The $\mu_{0}H_\mathrm{c2}$(0) data strongly correlate with the $T_\mathrm{s}$-dependent $T_\mathrm{c}$, indicating that higher $T_\mathrm{c}$ values lead to larger $\mu_{0}H_\mathrm{c2}$(0) in the thin films.

\begin{figure}
\begin{center}
\includegraphics[width=1\linewidth]{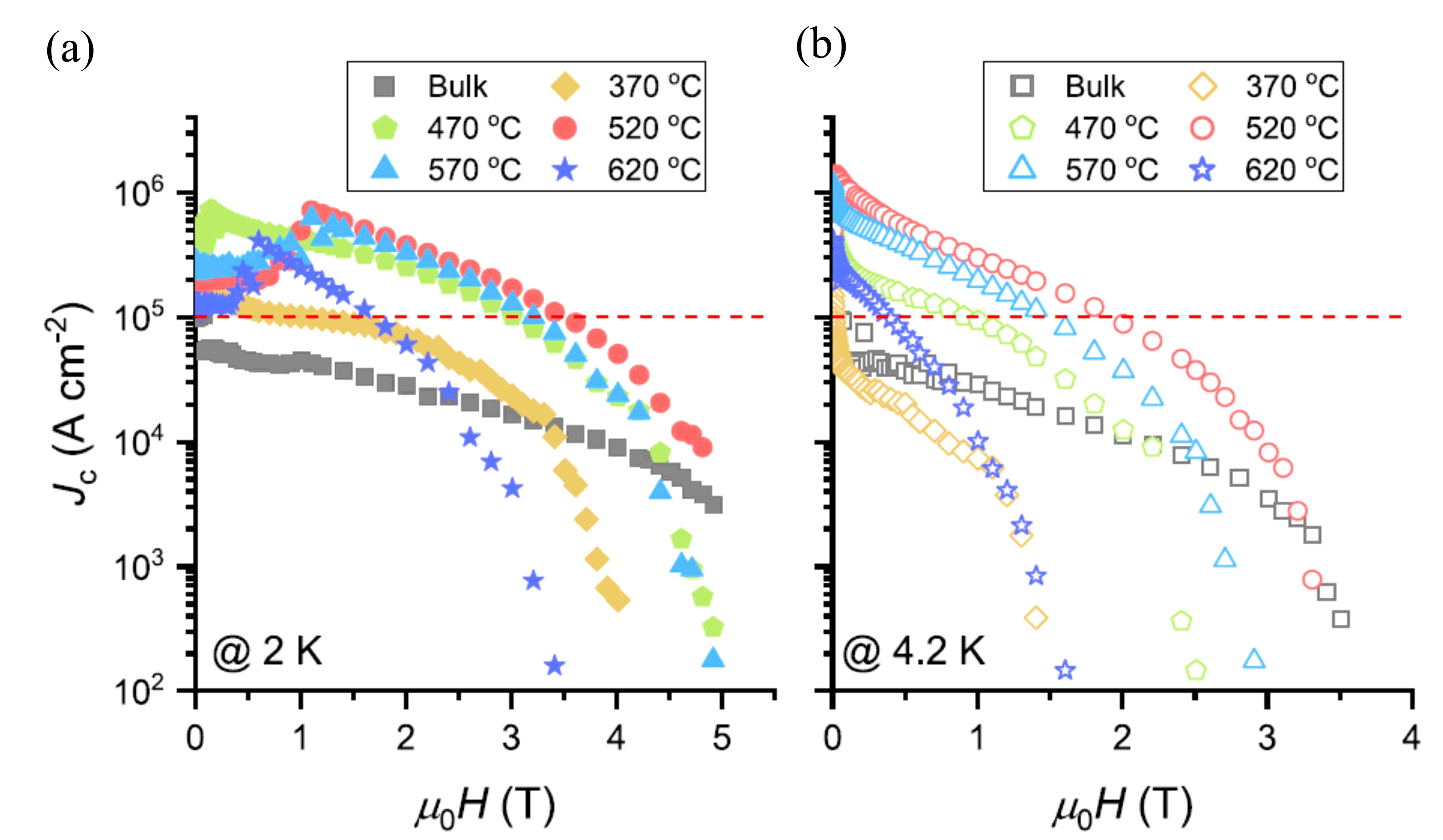}
\caption{\label{fig7} Magnetic field dependence of $J_\mathrm{c}$ for thin-film Ta$_{1/6}$Nb$_{2/6}$Hf$_{1/6}$Zr$_{1/6}$Ti$_{1/6}$ samples at (a) 2 K and (b) 4.2 K. The samples were fabricated using targets prepared through a ball-milling followed by a hot-press sintering process. Reproduced from \cite{Jung:NC2022} under Creative Commons Attribution License.}
\end{center}
\end{figure}

Figure\hspace{1mm}\ref{fig7} illustrates the magnetic field dependence of $J_\mathrm{c}$ for thin-film samples fabricated at various $T_\mathrm{s}$ values alongside data from the target sample (labeled as "Bulk"). 
A significant enhancement in $J_\mathrm{c}$ is observed for all thin-film samples compared to the bulk sample. 
Notably, the thin-film sample deposited at $T_\mathrm{s}$=520 $^{\circ}$C achieves $J_\mathrm{c}$ $>$ 1 MA/cm$^{2}$, representing approximately 820 \% and 790 \% increases relative to the bulk sample at 2.0 K (and 3.4 T) and 4.2 K (and 2 T), respectively. 
The red dashed line ($J_\mathrm{c}$=0.1 MA/cm$^{2}$) represents the benchmark for superconductors in large-scale applications, such as high-field superconducting magnets\cite{Larbalestier:Nature2001}. 
Most thin-film samples surpass this benchmark up to 4 T (at 2 K) and 2 T (at 4.2 K).
While the self-field $J_\mathrm{c}$ of thin-film superconductors typically exceeds that of bulk samples, the precise origin of the enhanced $J_\mathrm{c}$ in thin films remains unclear.

More recently, the Korean research group improved the low-field $J_\mathrm{c}$ of thin-film HEAs\cite{Jung:PSAC2024}. 
The team modified the target alloy preparation method, replacing the ball-milling and hot-press sintering process with arc melting followed by thermal annealing. 
Two samples, HEA550 and HEA700, were prepared, corresponding to films fabricated using targets annealed at 550 $^{\circ}$C and 700 $^{\circ}$C, respectively. 
Remarkably, HEA700 demonstrates a low-field $J_\mathrm{c}$ of approximately 4.4 MA/cm$^{2}$ (at 2 K) and 3.5 MA/cm$^{2}$ (at 4 K), outperforming the thin-film sample deposited at $T_\mathrm{s}$=520 $^{\circ}$C reported in Nature Communications.
The flux pinning force densities for the HEA550 and HEA700 samples are documented in Ref.\cite{Jung:PSAC2024}. 
In both samples, the force density at 2 K peaks at approximately 1 T, with maximum values of 2.34 GN/m$^{3}$ for HEA550 and 9.83 GN/m$^{3}$ for HEA700. 
The normal point pinning model describes the magnetic field dependence of flux pinning force density in each sample well.  
The authors attribute the point pinning mechanism to the severe lattice distortion effect in the "four core effects" of HEA.

The $J_\mathrm{c}$ values of the HEA thin-film samples reviewed here decline below 10$^{5}$ A/cm$^{2}$ as the magnetic field increases beyond approximately 3 T. 
The feasibility of their practical application as superconducting wires depends on the development of flux pinning sites effective at higher magnetic fields.

\subsection{Annealing effect on $J_\mathrm{c}$ of arc-melted sample}
\begin{figure}
\begin{center}
\includegraphics[width=0.8\linewidth]{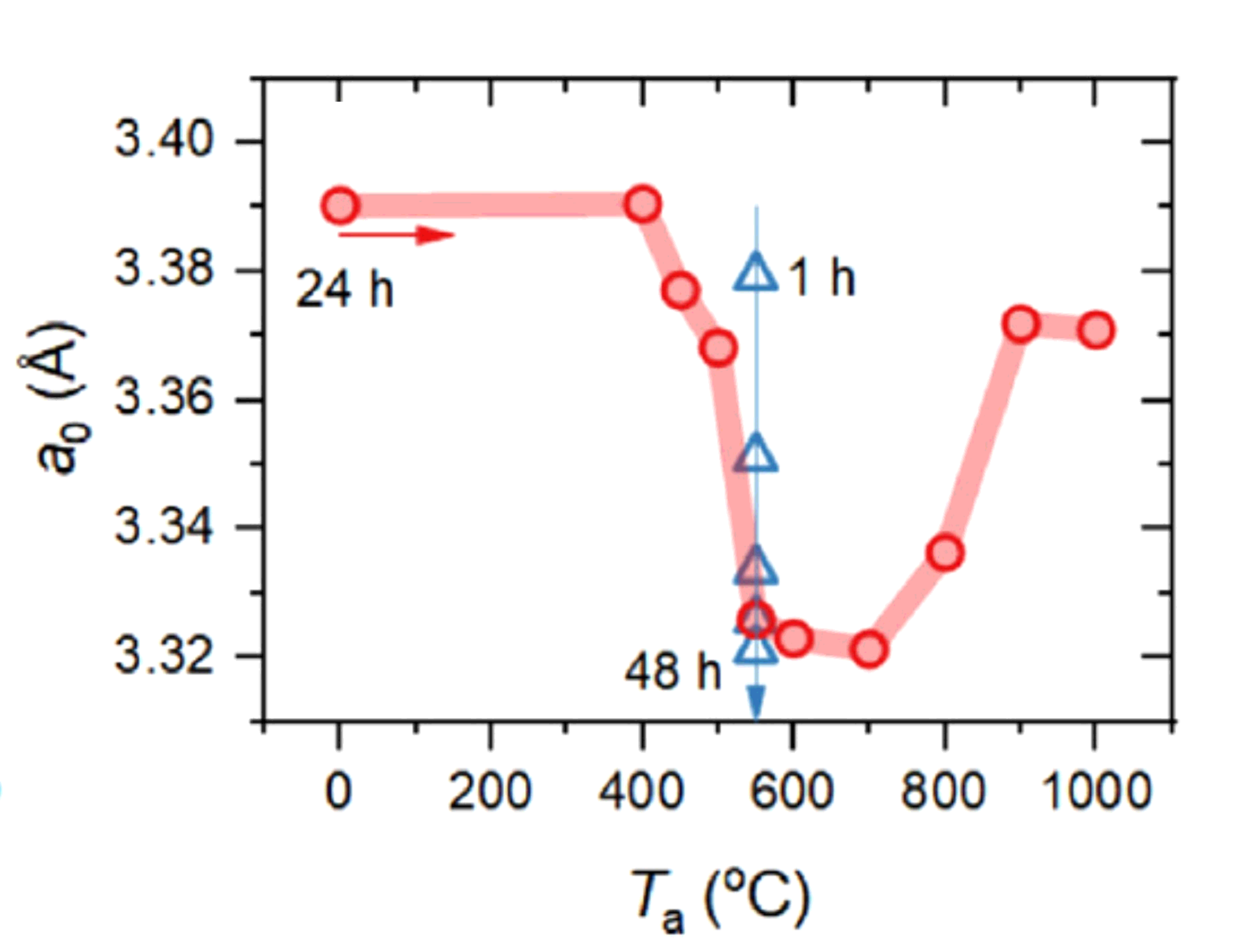}
\caption{\label{fig8} Dependence of lattice parameter on annealing temperature for Ta$_{1/6}$Nb$_{2/6}$Hf$_{1/6}$Zr$_{1/6}$Ti$_{1/6}$ synthesized via arc-melting technique. Reproduced with permission from \cite{Kim:JMST2024}.}
\end{center}
\end{figure}

In 2022, Gao et al. reported a remarkable enhancement of $J_\mathrm{c}$ in thermally annealed bulk (TaNb)$_{0.7}$(HfZrTi)$_{0.5}$  HEA, as discussed below\cite{Gao:APL2022}. 
Similarly, in 2024, the Korean research team investigated the annealing effects in bulk Ta$_{1/6}$Nb$_{2/6}$Hf$_{1/6}$Zr$_{1/6}$Ti$_{1/6}$, reporting a thermally-driven dramatic enhancement of $J_\mathrm{c}$\cite{Kim:JMST2024}. 
The arc-melted samples were thermally annealed for 24 hours at temperatures ranging from 400 $^{\circ}$C to 1000 $^{\circ}$C.
Notably, the lattice parameter $a_{0}$ demonstrated significant sensitivity to the annealing temperature (Fig.\hspace{1mm}\ref{fig8}). 
While $a_{0}$ remained nearly constant up to 400 $^{\circ}$C, it began to decrease sharply above this threshold, reaching a minimum of approximately 3.32 \AA \hspace{0.5mm} at 700 $^{\circ}$C, before increasing again with higher annealing temperatures. 
The contraction of $a_{0}$ to 3.32 \AA \hspace{0.5mm} suggests the introduction of lattice strain in the annealed Ta$_{1/6}$Nb$_{2/6}$Hf$_{1/6}$Zr$_{1/6}$Ti$_{1/6}$.

The annealing temperature dependence of $T_\mathrm{c}$ exhibits only minor variation with annealing temperature. 
Conversely, $\mu_{0}H_\mathrm{c2}$(0) decreases monotonically with increasing annealing temperature. 
The value of $\mu_{0}H_\mathrm{c2}$(0), initially 13.27 T in the as-cast sample, diminishes to 9.07 T after annealing at 1000 $^{\circ}$C, a trend that does not strongly correlate with the annealing effects on $T_\mathrm{c}$.

\begin{figure}
\begin{center}
\includegraphics[width=1\linewidth]{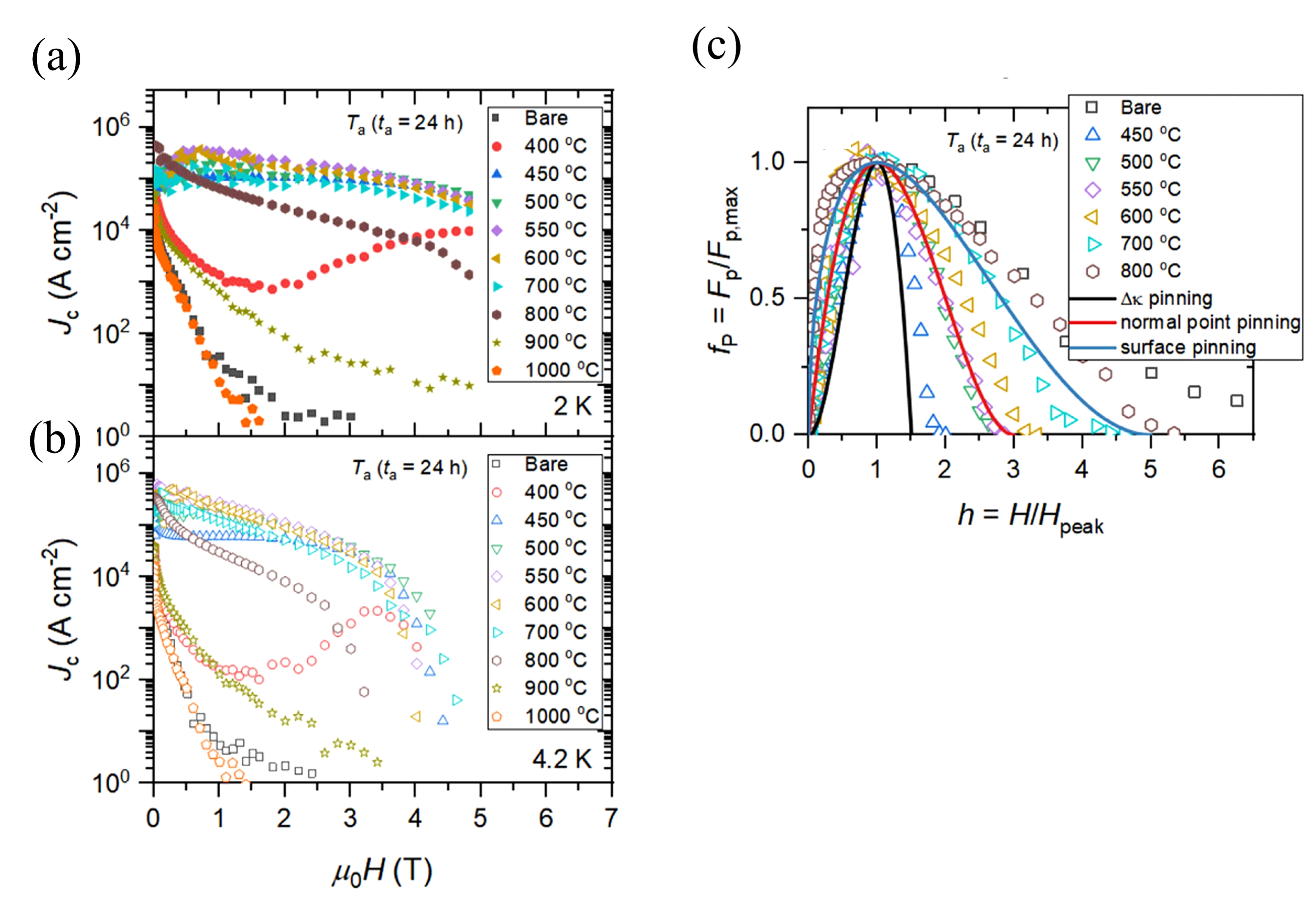}
\caption{\label{fig9} (a) Field-dependent $J_\mathrm{c}$ at 2 K, (b) field-dependent $J_\mathrm{c}$ at 4.2 K, and (c) normalized flux pinning force density $f_\mathrm{p}$ as a function of the reduced field $h$ for heat-treated  
Ta$_{1/6}$Nb$_{2/6}$Hf$_{1/6}$Zr$_{1/6}$Ti$_{1/6}$. Reproduced with permission from \cite{Kim:JMST2024}.}
\end{center}
\end{figure}

Thermal annealing induces more significant changes in $J_\mathrm{c}$ than in $T_\mathrm{c}$ or $\mu_{0}H_\mathrm{c2}$(0) (Figs.\hspace{1mm}\ref{fig9}(a) and (b)). 
In these figures, the "Bare sample" refers to the as-cast state. 
The self-field $J_\mathrm{c}$ of the bare sample is approximately 39 kA/cm$^{2}$ at 2 K and 28 kA/cm$^{2}$ at 4.2 K, both decreasing rapidly under an external magnetic field. 
However, samples annealed at 500 - 700 $^{\circ}$C exhibit substantial improvements in self-field $J_\mathrm{c}$ and field performance. 
For instance, the self-field $J_\mathrm{c}$ at 4.2 K for the sample annealed at 550 $^{\circ}$C is enhanced by approximately 1860 \% compared to the bare sample. 
In terms of field performance, the $J_\mathrm{c}$ at 2 K for the 550 $^{\circ}$C annealed sample surpasses 100 kA/cm$^{2}$ even at 4 T, whereas the bare sample's $J_\mathrm{c}$ drops to 10 A/cm$^{2}$ under a field of $\sim$1.5 T. 
Overall behaviors depicted in Figs.\hspace{1mm}\ref{fig9}(a) and (b) suggest that the 550 $^{\circ}$C annealed sample exhibits the best $J_\mathrm{c}$ performance.
The flux pinning force densities, $F_\mathrm{p}$, at 4.2 K are compared among samples subjected to different annealing conditions. 
The highest $F_\mathrm{p}$ value of 3 GN/m$^{3}$ is achieved in the sample annealed at 550 $^{\circ}$C.
A systematic shift in the magnetic field corresponding to the maximum $F_\mathrm{p}$ is observed, ranging from approximately 2.5 T in the 450 $^{\circ}$C annealed sample to 0.1 T in the 800 $^{\circ}$C annealed sample. 
This shift suggests a possible crossover in the pinning mechanism with increasing annealing temperature.
Fig.\hspace{1mm}\ref{fig9}(c) presents an analysis of data from various samples using three pinning models. 
While the pinning mechanism strongly depends on the annealing temperature, the $f_\mathrm{p}$ behavior of the 550 $^{\circ}$C annealed sample aligns with the normal point pinning model. 
Considering the pronounced lattice strain observed in the 550 $^{\circ}$C annealed sample, it is inferred that this strain acts as a point-pinning center, effectively impeding flux motion under an applied magnetic field.

\subsection{Fabrication and characterization of wires}\label{wire}
In 2024, the Korean team reported the fabrication and $J_\mathrm{c}$ evaluation of Ta$_{1/6}$Nb$_{2/6}$Hf$_{1/6}$Zr$_{1/6}$Ti$_{1/6}$ wires\cite{Jung:JALCOM2024}. 
This study holds significant value in assessing the feasibility of HEA-based superconducting wires. Ta$_{1/6}$Nb$_{2/6}$Hf$_{1/6}$Zr$_{1/6}$Ti$_{1/6}$ wires were produced using the ex-situ powder-in-tube (PIT) method. 
The precursor powder, comprising a stoichiometric mixture of elemental powders, was prepared via high-energy planetary ball milling. 
The prepared precursor was loaded into Fe metal tubes, which were then drawn into wires. 
The wires, cut into 30 mm lengths, were sintered in an evacuated quartz tube for 1 hour at various sintering temperatures ($T_\mathrm{s}$=600 $\sim$ 1000 $^{\circ}$C).

\begin{figure}
\begin{center}
\includegraphics[width=0.8\linewidth]{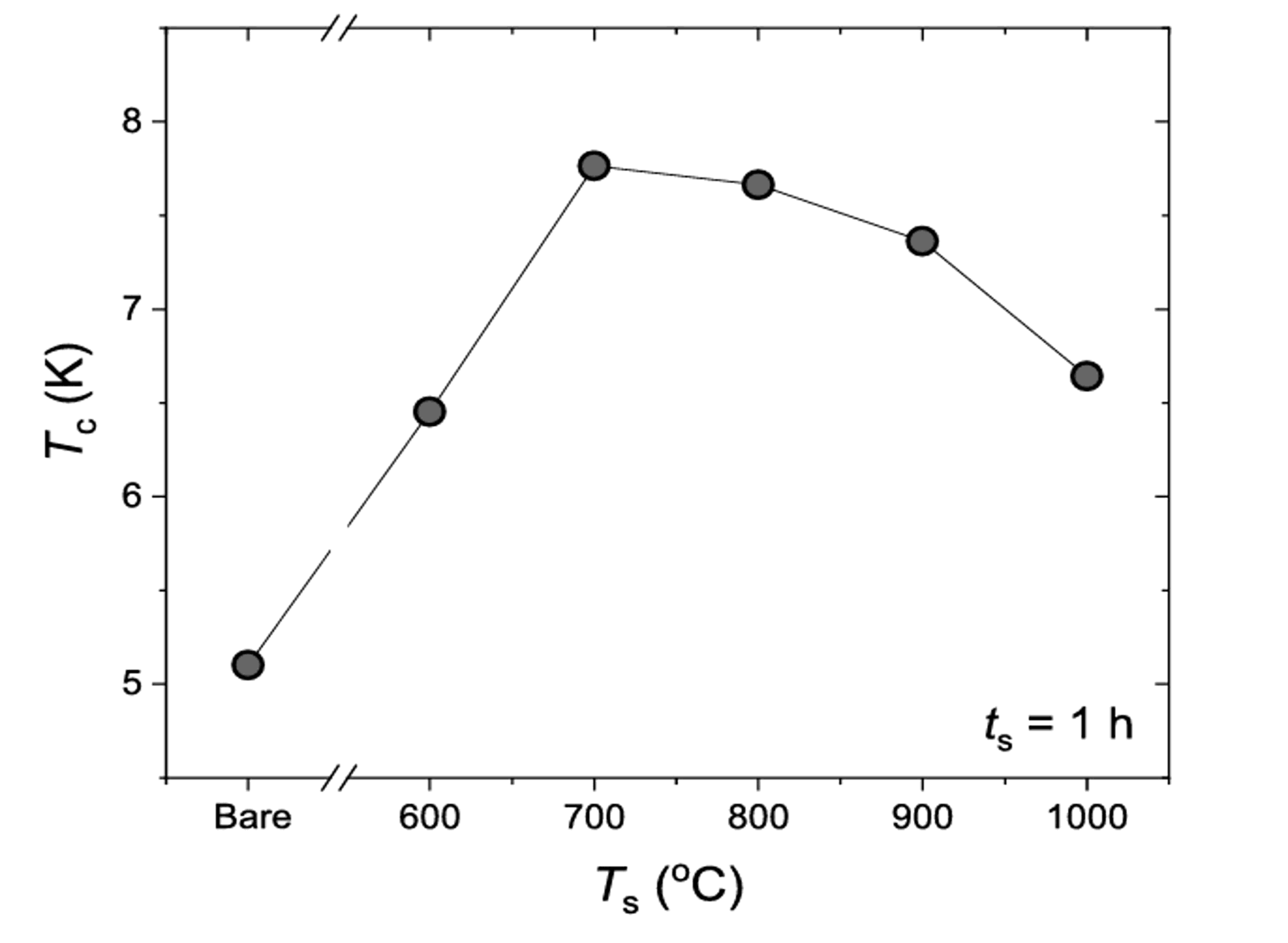}
\caption{\label{fig10} Sintering temperature dependence of $T_\mathrm{c}$ for Ta$_{1/6}$Nb$_{2/6}$Hf$_{1/6}$Zr$_{1/6}$Ti$_{1/6}$ wire. Reproduced with permission from \cite{Jung:JALCOM2024}.}
\end{center}
\end{figure}

In Fig.\hspace{1mm}\ref{fig10}, $T_\mathrm{c}$ is plotted as a function of $T_\mathrm{s}$. 
$T_\mathrm{c}$ increases markedly from 5.10 K in the untreated wire (bare state) to 7.76 K in the wire sintered at 700 $^{\circ}$C. 
However, further increases in $T_\mathrm{s}$ lead to a gradual decrease in $T_\mathrm{c}$. 
The authors attribute the enhanced $T_\mathrm{c}$ at $T_\mathrm{s}$ between 500 and 700 $^{\circ}$C to the formation of a Ta/Nb-rich bcc phase. 
Conversely, the decline in $T_\mathrm{c}$ above 700 $^{\circ}$C is speculated to result from the emergence of impurity phases, such as Hf$_{2}$Fe and Zr$_{2}$Fe, with the Fe contamination arising from the Fe tube used in the PIT process. 
Unlike the $T_\mathrm{c}$ trends, $\mu_{0}H_\mathrm{c2}$(0) for wires sintered between 600 and 900 $^{\circ}$C exhibits minimal variation, maintaining a high value of 10.3 T, corresponding to a coherence length of 5.66 nm. 
This stability of $\mu_{0}H_\mathrm{c2}$(0) across different annealing conditions is advantageous for practical superconducting wire applications.

\begin{figure}
\begin{center}
\includegraphics[width=1\linewidth]{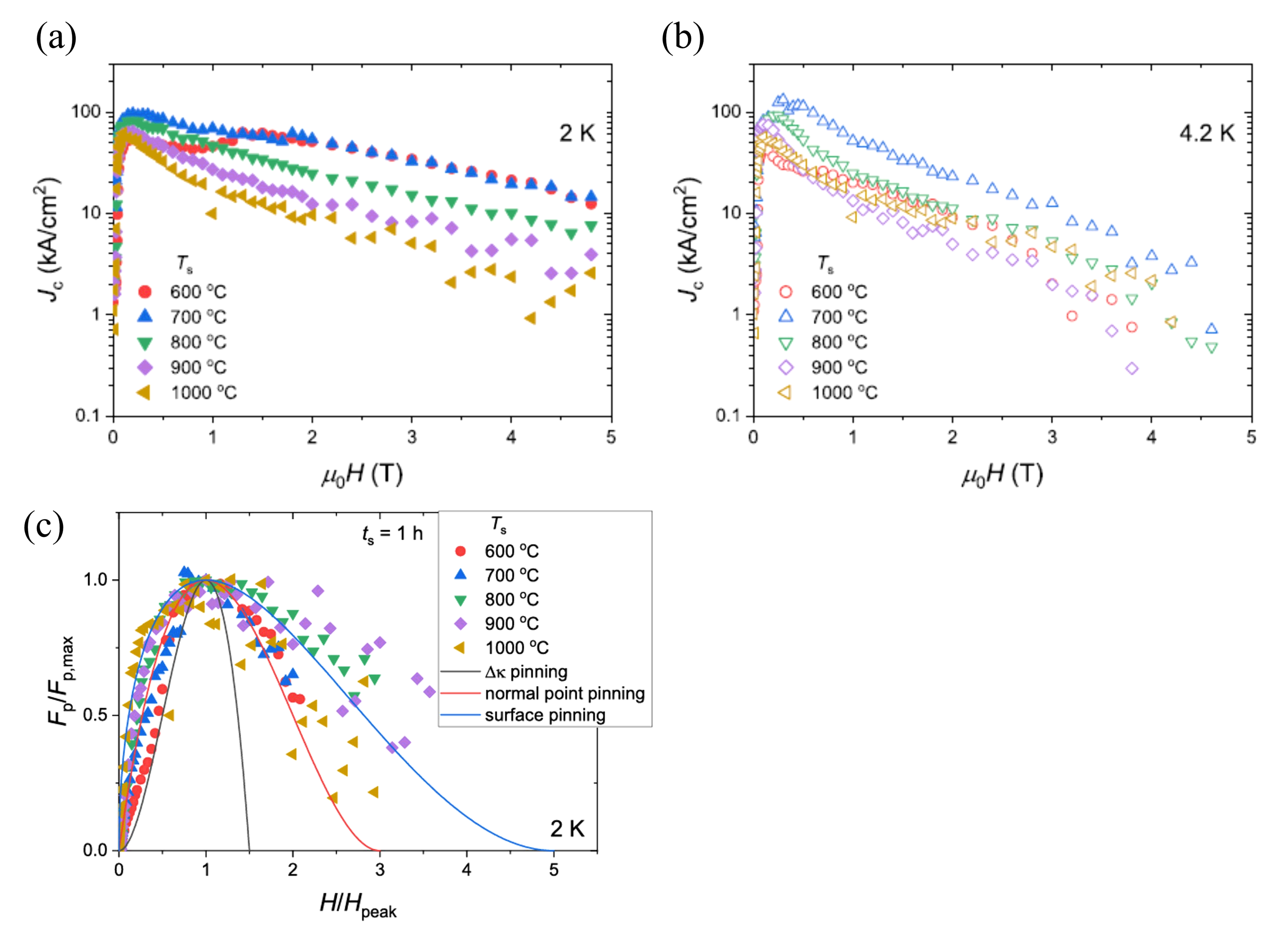}
\caption{\label{fig11} Field-dependent $J_\mathrm{c}$ of Ta$_{1/6}$Nb$_{2/6}$Hf$_{1/6}$Zr$_{1/6}$Ti$_{1/6}$ wire at (a) 2 K and (b) 4.2 K. (c) Normalized flux pinning force density as a function of $H/H_\mathrm{peak}$ for Ta$_{1/6}$Nb$_{2/6}$Hf$_{1/6}$Zr$_{1/6}$Ti$_{1/6}$ wire. Reproduced with permission from \cite{Jung:JALCOM2024}.}
\end{center}
\end{figure}

Figures \hspace{1mm}\ref{fig11}(a) and (b) present the magnetic field-dependent $J_\mathrm{c}$ for each wire. 
The low-field $J_\mathrm{c}$ at 4.2 K for the wire sintered at 700 $^{\circ}$C exceeds 100 kA/cm$^{2}$, meeting the threshold for practical applications. 
However, further enhancement of $J_\mathrm{c}$ under higher magnetic fields remains desirable. 
Similar to heat-treated Ta$_{1/6}$Nb$_{2/6}$Hf$_{1/6}$Zr$_{1/6}$Ti$_{1/6}$, a systematic decrease in the magnetic field corresponding to the maximum flux pinning force density is observed with increasing sintering temperature from 600 $^{\circ}$C to 1000 $^{\circ}$C. 
The sample sintered at 700 $^{\circ}$C exhibits the highest pinning force density of 1.1 GN/m$^{3}$.
The analysis of the pinning mechanism, shown in Fig.\hspace{1mm}\ref{fig11}(c), indicates that most wires are dominated by normal point pinning. 
This finding aligns with observations in thermally annealed Ta$_{1/6}$Nb$_{2/6}$Hf$_{1/6}$Zr$_{1/6}$Ti$_{1/6}$ bulk samples\cite{Kim:JMST2024}. 
Hence, lattice strain likely plays a pivotal role in magnetic flux pinning. 
Additionally, impurity phases such as Hf$_{2}$Fe and Zr$_{2}$Fe may serve as alternative sources of point pinning.

Some wire samples exhibit $J_\mathrm{c}$ exceeding 10$^{5}$ A/cm$^{2}$ at lower magnetic fields; however, further efforts are necessary to enhance $J_\mathrm{c}$, particularly at higher fields. 
While the sintering method is currently the sole approach employed for wire fabrication, exploring alternative fabrication techniques is highly desirable.
For instance, drawing a rod-shaped sample obtained via arc melting represents a promising alternative method.

\begin{figure}
\begin{center}
\includegraphics[width=1\linewidth]{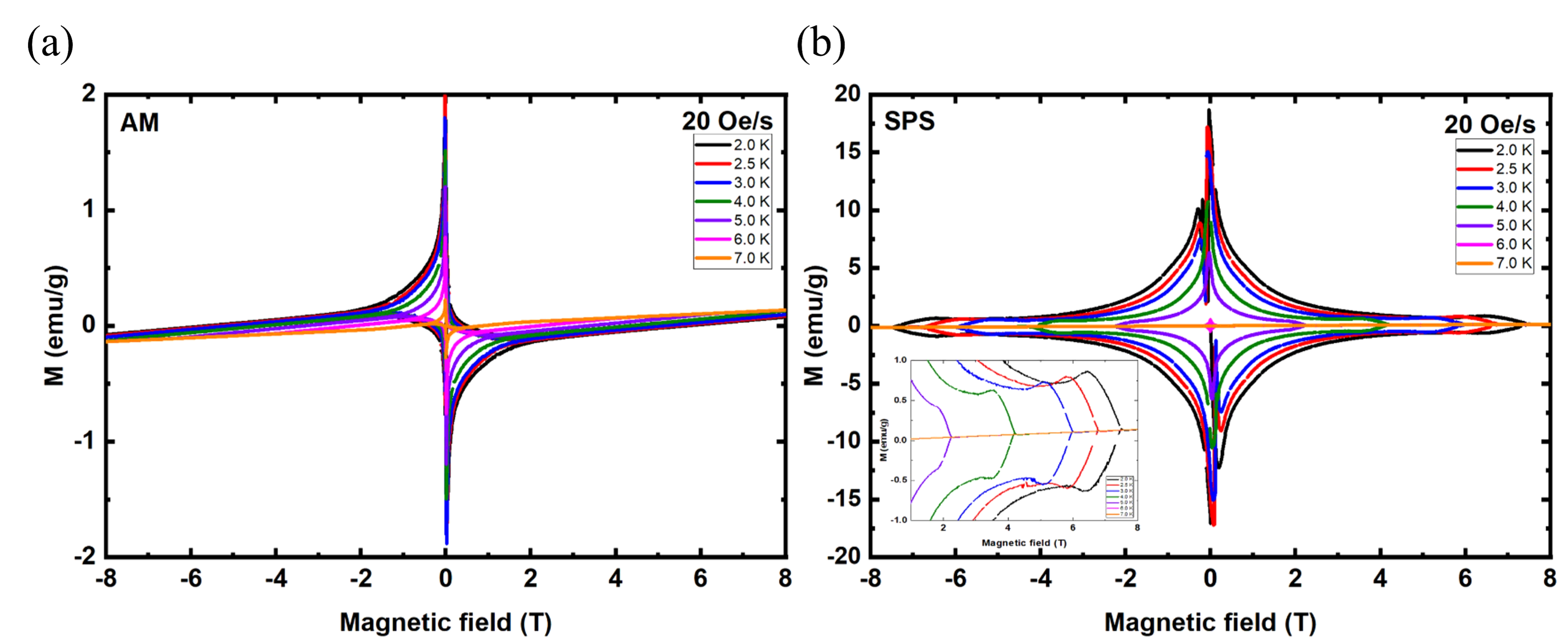}
\caption{\label{fig12} Isothermal magnetic hysteresis loops of Nb$_{2/5}$Hf$_{1/5}$Zr$_{1/5}$Ti$_{1/5}$ for (a) arc-melted sample and (b) sample synthesized via ball milling and spark plasma sintering (SPS) methods. Reproduced with permission from \cite{Hidayati:AM2023}.}
\end{center}
\end{figure}

\subsection{Nb$_{2/5}$Hf$_{1/5}$Zr$_{1/5}$Ti$_{1/5}$}
The Korean team investigated the effect of removing Ta from Ta$_{1/6}$Nb$_{2/6}$Hf$_{1/6}$Zr$_{1/6}$Ti$_{1/6}$ and evaluated $J_\mathrm{c}$ of the quaternary HEA Nb$_{2/5}$Hf$_{1/5}$Zr$_{1/5}$Ti$_{1/5}$, synthesized using arc-melting and spark plasma sintering (SPS) techniques\cite{Hidayati:AM2023}. 
The superconducting transition temperatures were 7.99 K for the arc-melted sample and 6.63 K for the SPS sample. 
The upper critical fields were 14.63 T and 11.68 T for the arc-melted and SPS samples, respectively, corresponding to $\xi_\mathrm{GL}$(0) of 4.84 nm and 5.41 nm.

\begin{figure}
\begin{center}
\includegraphics[width=1\linewidth]{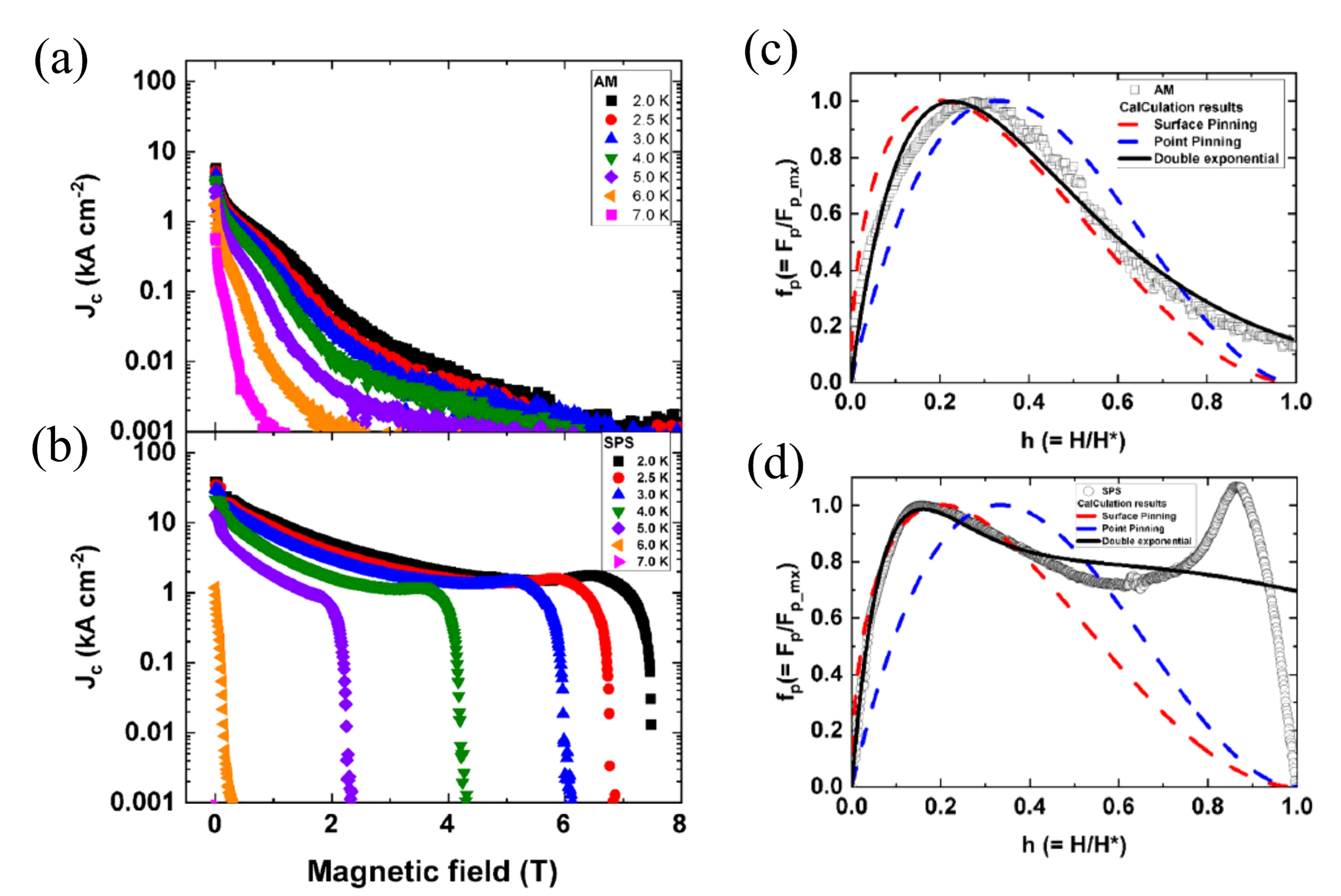}
\caption{\label{fig13} Field-dependent $J_\mathrm{c}$ of Nb$_{2/5}$Hf$_{1/5}$Zr$_{1/5}$Ti$_{1/5}$ for (a) arc-melted sample and (b) sample synthesized via ball-milling and spark plasma sintering (SPS) methods. Normalized flux pinning force density $f_\mathrm{p}$ as a function of the reduced field $h$ for Nb$_{2/5}$Hf$_{1/5}$Zr$_{1/5}$Ti$_{1/5}$ for (c) arc-melted sample and (d) SPS sample. Reproduced with permission from \cite{Hidayati:AM2023}.}
\end{center}
\end{figure}

The magnetic hysteresis loops of the arc-melted sample exhibit narrow and asymmetric profiles (see Fig.\hspace{1mm}\ref{fig12}(a)). 
The asymmetry in the $M$-$H$ loops has been attributed to the surface barrier effect\cite{Bean:PRL1964}.
Conversely, the SPS sample displays symmetric, diamond-shaped hysteresis loops, indicating strong flux pinning. 
Flux jumps are observed at low temperatures and low magnetic fields in the SPS sample (see Fig.\hspace{1mm}\ref{fig12}(b)). 
Furthermore, at higher magnetic fields, the $M$-$H$ loops of the SPS sample show an unusual secondary peak, known as the fishtail effect, suggesting the presence of flux pinning sites that become effective at elevated magnetic fields (see the inset of Fig.\hspace{1mm}\ref{fig12}(b)).
Figures\hspace{1mm}\ref{fig13}(a) and (b) present the $J_\mathrm{c}$ datasets for the arc-melted and SPS samples. 
Notably, the self-field $J_\mathrm{c}$ at 2 K for the SPS sample (39,000 A/cm$^{2}$) is significantly higher than that of the arc-melted sample (5,500 A/cm$^{2}$). 
Moreover, the SPS sample demonstrates superior performance under external magnetic fields, with a more gradual decrease in $J_\mathrm{c}$, likely attributed to the fishtail effect. 
As discussed in subsection \ref{subsecSPS}, Fe contamination from the ball-milling process during SPS preparation introduces magnetic flux pinning, contributing to the enhanced $J_\mathrm{c}$ of the SPS sample.
Figures\hspace{1mm}\ref{fig13}(c) and (d) compare the experimental $f_\mathrm{p}(h)$ of the arc-melted and SPS samples against several pinning models. 
In the arc-melted sample, both surface and point pinning models align well with the experimental data. 
For the SPS sample, $f_\mathrm{p}(h)$ at lower fields corresponds to surface pinning or the double exponential model. 
However, deviations are observed at higher magnetic fields due to the fishtail effect, which drives an unusual increase in $f_\mathrm{p}(h)$.

\begin{figure}
\begin{center}
\includegraphics[width=1.0\linewidth]{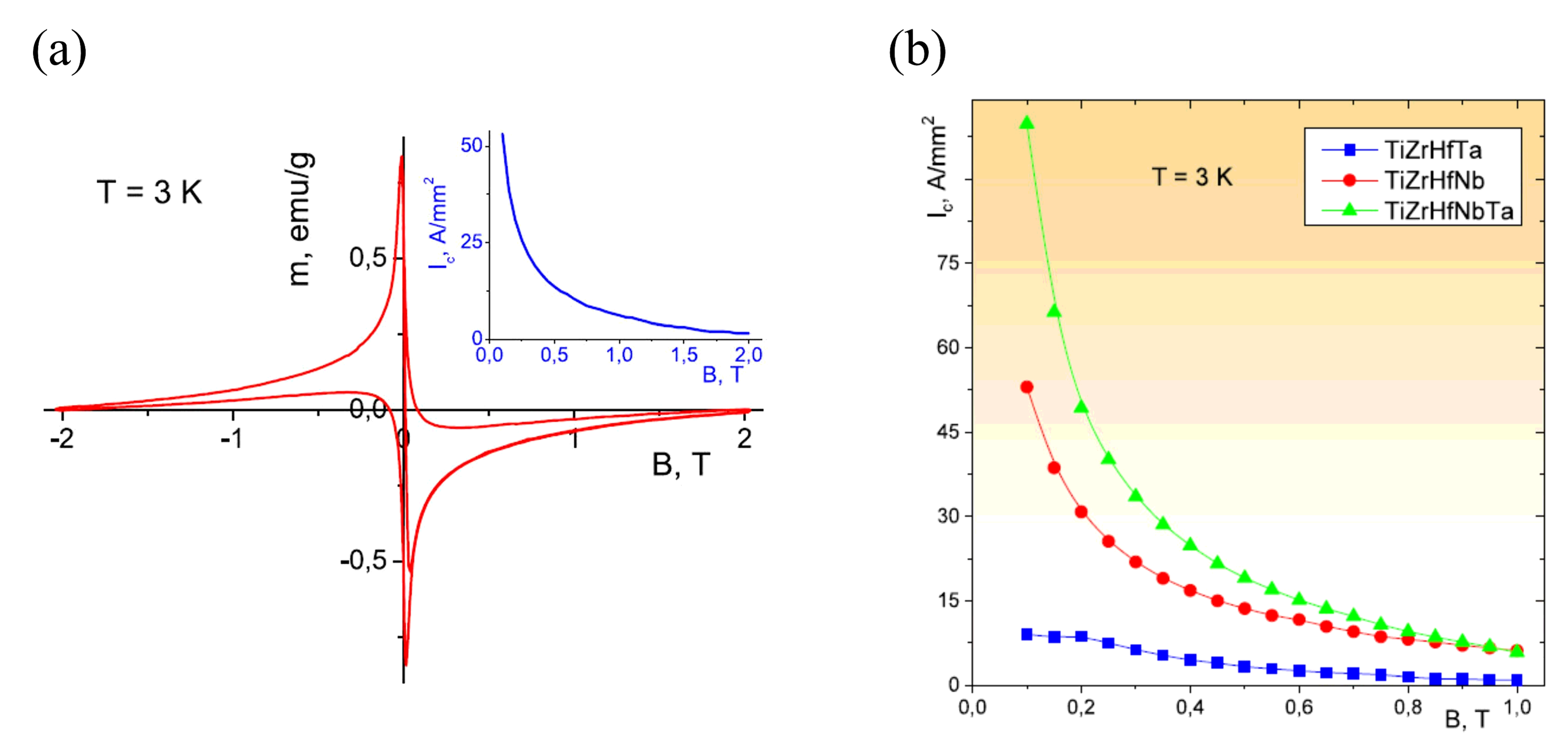}
\caption{\label{fig14} (a) Isothermal magnetization curve and magnetic field dependence of $J_\mathrm{c}$ for TiZrHfNb at 3 K. Reproduced with permission from \cite{Uporpv:IM2022}. (b) Magnetic field dependence of $J_\mathrm{c}$ for TiZrHfTa, TiZrHfNb, and TiZrHfNbTa. Reproduced with permission from \cite{Uporpv:IM2025}.}
\end{center}
\end{figure}

\subsection{TiZrHfNb, TiZrHfTa, TiZrHfNbTa, and Ti-V-Nb-Ta}
Uporov et al. reported the $J_\mathrm{c}$ performance of equimolar TiZrHfNb\cite{Uporpv:IM2022}, prepared using the arc-melting method and exhibiting a single-phase structure. 
The $T_\mathrm{c}$ and upper critical field were determined as 6.3 K and 8 T, respectively. 
The estimated $J_\mathrm{c}$ at 3 K is moderate compared to superconductors based on Nb–Ti alloys (see Fig.\hspace{1mm}\ref{fig14}(a)). 
Recently, Uporov et al. extended their $J_\mathrm{c}$ investigations to TiZrHfTa and TiZrHfNbTa (see Fig.\hspace{1mm}\ref{fig14}(b))\cite{Uporpv:IM2025}.
However, the $J_\mathrm{c}$ values of these alloys remain below the practical threshold of 10$^{5}$ A/cm$^{2}$.

Sharma et al. investigated the $J_\mathrm{c}$ performance of the bcc-structured (TiV)$_{0.5}$Nb$_{0.4}$Ta$_{0.1}$ alloy, which exhibits $T_\mathrm{c}$=5.2 K and $\mu_{0}H_\mathrm{c2}(0)$=6.4 T\cite{Sharma:MTC2025}. 
The magnetic field dependence of $J_\mathrm{c}$ at 1.8 K has been reported, with the self-field $J_\mathrm{c}$ reaching 0.4$\times$10$^{5}$ A/cm$^{2}$, monotonically decreasing as the magnetic field increases. 
The dominant flux pinning mechanism in this alloy is surface pinning.

\begin{figure}
\begin{center}
\includegraphics[width=1\linewidth]{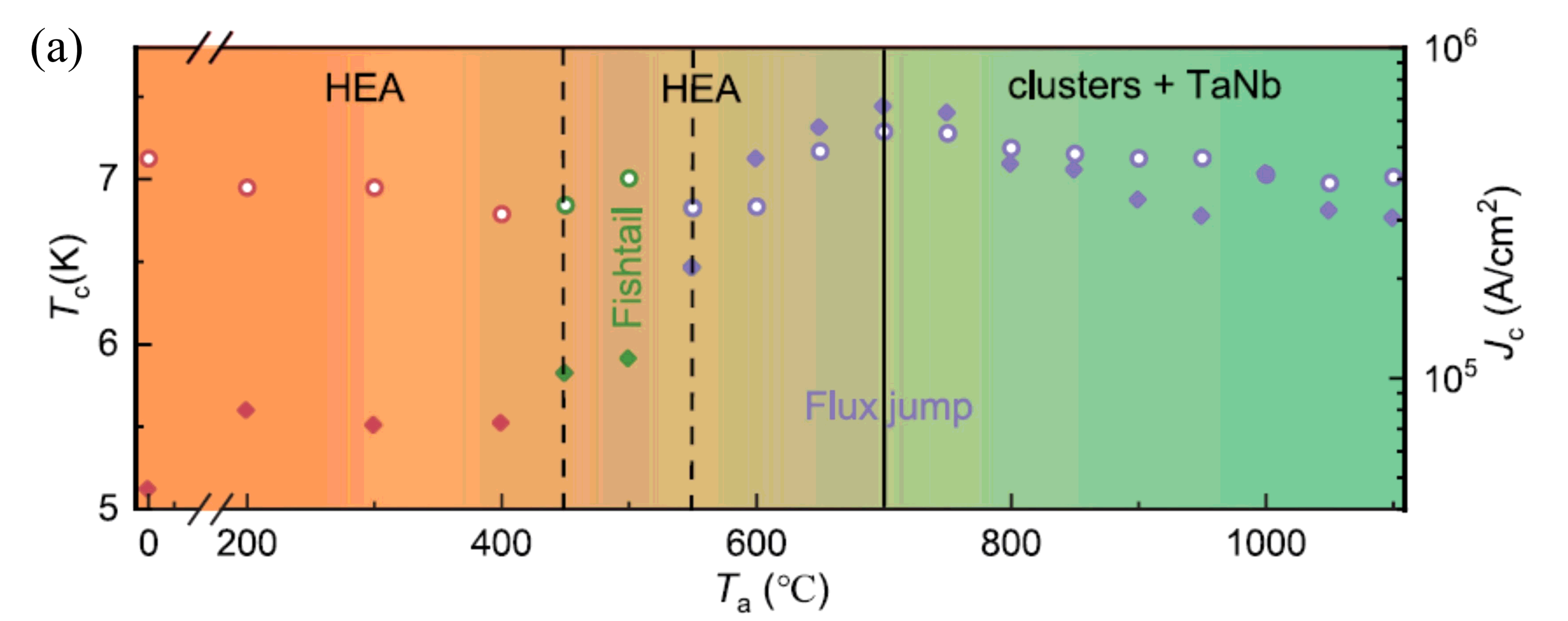}
\caption{\label{fig15} Dependence of $T_\mathrm{c}$ (open circles) and maximum $J_\mathrm{c}$ at 2 K (filled diamonds) on annealing temperature for (TaNb)$_{0.7}$(HfZrTi)$_{0.5}$. Reproduced with permission from \cite{Gao:APL2022}.}
\end{center}
\end{figure}
 
\section{(TaNb)$_{0.7}$(HfZrTi)$_{0.5}$}\label{sec4}
(TaNb)$_{0.7}$(HfZrTi)$_{0.5}$ can be categorized as a Ta-Nb-Hf-Zr-Ti Senkov alloy, as discussed in the previous chapter. 
However, its $J_\mathrm{c}$ dependence on annealing temperature and a fishtail effect are distinctive features. 
Therefore, we address this alloy in a separate chapter.

To the best of our knowledge, the annealing effect on $J_\mathrm{c}$ in HEAs has been investigated for the first time in this particular alloy by Gao et al\cite{Gao:APL2022}. 
A polycrystalline sample was prepared using the arc-melting method and subsequently annealed at various temperatures ($T_\mathrm{a}$) for 5 hours. 
Figure\hspace{1mm}\ref{fig15} presents the phase diagram of (TaNb)$_{0.7}$(HfZrTi)$_{0.5}$, revealing that the bcc structure remains stable up to $T_\mathrm{a}$=500 $^{\circ}$C. 
At $T_\mathrm{a}$=700 $^{\circ}$C, the Bragg reflection peaks in the X-ray diffraction (XRD) pattern shift to higher angles, indicating a reduction in the lattice parameter. 
The authors attribute this phenomenon to the emergence of TaNb with a bcc structure.
Furthermore, metallographic analysis of the sample annealed at 500 $^{\circ}$C revealed the precipitation of clusters approximately 10 nm in size. 
$T_\mathrm{c}$=7.13 K of the as-cast sample remains nearly unchanged after annealing. 
In the sample annealed at 500 $^{\circ}$C,  $\mu_{0}H_\mathrm{c2}$(0) is estimated to be 10.07 T, corresponding to $\xi_\mathrm{GL}$(0) of 5.72 nm.

\begin{figure}
\begin{center}
\includegraphics[width=1\linewidth]{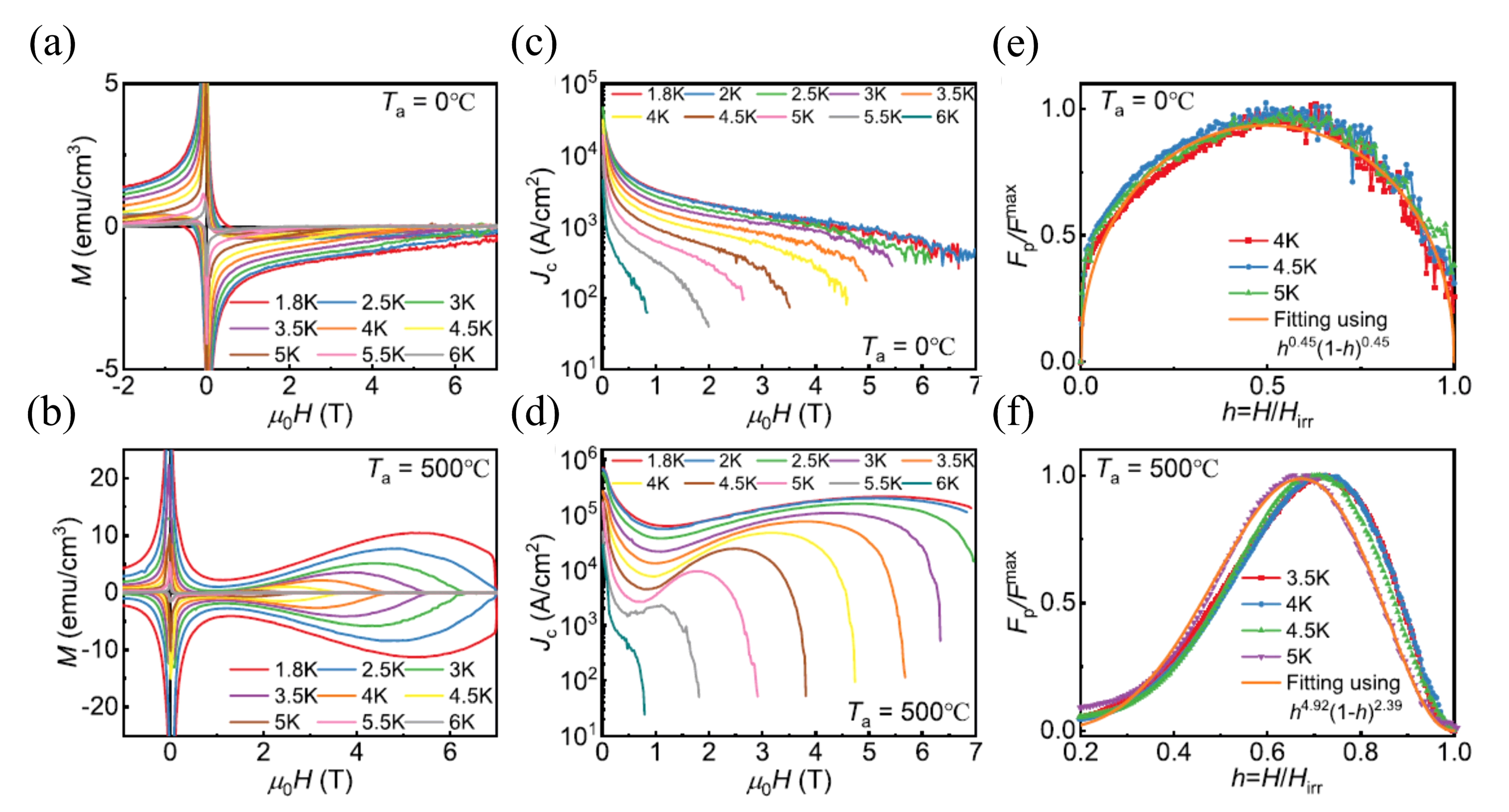}
\caption{\label{fig16} (a) and (b) Isothermal magnetization curves of (TaNb)$_{0.7}$(HfZrTi)$_{0.5}$ for as-cast and 500 $^{\circ}$C annealed samples, respectively. (c) and (d) Magnetic field-dependent $J_\mathrm{c}$ of (TaNb)$_{0.7}$(HfZrTi)$_{0.5}$ for as-cast and 500 $^{\circ}$C annealed samples, respectively. (e) and (f) Normalized flux pinning force density as a function of reduced field $h$ for (TaNb)$_{0.7}$(HfZrTi)$_{0.5}$ for as-cast and 500 $^{\circ}$C annealed samples, respectively. Reproduced with permission from \cite{Gao:APL2022}.}
\end{center}
\end{figure}

Figures\hspace{1mm}\ref{fig16}(a)-(f) summarize the $M$-$H$ curves, magnetic field dependencies of $J_\mathrm{c}$, and $F_\mathrm{p}/F_\mathrm{p,max}$ curves for the as-cast and 500 $^{\circ}$C-annealed samples. 
While the as-cast sample exhibits magnetic hysteresis loops characteristic of a conventional alloy superconductor, a fishtail effect is observed in the 500 $^{\circ}$C-annealed sample at high magnetic fields. 
The second peak in the magnetization shifts to higher fields and becomes more pronounced as the temperature decreases. 
The self-field $J_\mathrm{c}$ of the sample annealed at 500 $^{\circ}$C is 6.81$\times$10$^{5}$ A/cm$^{2}$ at 2 K, an order of magnitude higher than that of the as-cast sample. 
Notably, the $J_\mathrm{c}$ of the annealed sample exhibits a broad maximum in the high-field region due to the fishtail effect. 
Similar behavior is observed in YBCO and Ba$_{0.6}$K$_{0.4}$Fe$_{2}$As$_{2}$, where the effect originates from the small-size normal core pinning\cite{Daeumling:Nature1990,Yang:APL2008}.
Significantly, the high-field $J_\mathrm{c}$ at 2 K in the annealed sample surpasses the practical level 10$^{5}$ A/cm$^{2}$.
The symmetric $F_\mathrm{p}/F_\mathrm{p,max}$ curves of the as-cast sample are well-fitted by the function $h^{0.45}$(1-$h$)$^{0.45}$.
In contrast, for the sample annealed at 500 $^{\circ}$C, the curves are better described by $h^{4.92}$(1-$h$)$^{2.39}$, contrasting sharply with high-$T_\mathrm{c}$ superconductors such as YBCO and Ba$_{0.6}$K$_{0.4}$Fe$_{2}$As$_{2}$\cite{Daeumling:Nature1990,Yang:APL2008}. 
This suggests that the present HEA possesses multiple pinning mechanisms. 
The reduction in lattice parameter during annealing supports the presence of lattice strain akin to that observed in Ta$_{1/6}$Nb$_{2/6}$Hf$_{1/6}$Zr$_{1/6}$Ti$_{1/6}$, which likely enhances $J_\mathrm{c}$, particularly at lower magnetic fields.
In (TaNb)$_{0.7}$(HfZrTi)$_{0.5}$, clusters precipitated from the HEA, measuring approximately 10 nm, are also observed. 
This nanostructure is reminiscent of $\alpha$-Ti precipitates in Nb-Ti and may play a crucial role in the fishtail effect, contributing to the improved high-field magnetic performance of $J_\mathrm{c}$. 
Thus, the lattice parameter and microstructure of thermally annealed (TaNb)$_{0.7}$(HfZrTi)$_{0.5}$ provide valuable insights for its potential application as a superconducting wire.

\section{NbScTiZr}\label{sec5}
NbScTiZr exhibits an eutectic microstructure, distinguishing it from conventional Nb-Ti alloys. 
Notably, the $T_\mathrm{c}$ of NbScTiZr is relatively higher than those of typical HEA superconductors, which can be partially attributed to the lattice strain induced by its fine eutectic structure. 
In practical superconducting wires, a multifilamentary structure is employed to mitigate the thermal instability of the superconducting state. 
The fine eutectic microstructure can be considered an inherent multifilamentary structure, representing one of the distinctive features of eutectic HEAs.

\subsection{Eutectic microstructure}
A eutectic alloy is defined as a homogeneous mixture with a melting point lower than those of its constituent phases. 
During solidification, two distinct solid phases emerge simultaneously, forming a characteristic microstructure, such as a lamellar arrangement. 
Eutectic HEAs have garnered significant attention as structural materials due to their potential to overcome the conventional strength-ductility trade-off\cite{Lu:SM2020,Bhardwaj:TI2021}. 
The fcc HEAs typically exhibit high ductility but low strength, whereas many bcc HEAs demonstrate high strength but poor ductility. 
Consequently, eutectic HEAs comprising both fcc and bcc phases achieve a balance of high strength and ductility.

In superconductors, the presence of eutectic phases often modifies the superconducting properties. 
For instance, an enhancement in $T_\mathrm{c}$ due to eutectic microstructures has been reported in materials such as Sr$_{2}$RuO$_{4}$, YIr$_{2}$, and Zr$_{5}$Pt$_{3}$O$_{x}$\cite{Maeno:PRL1998,Matthias:Science1980,Hamamoto:MRX2018}.
However, the influence of eutectic microstructures on $J_\mathrm{c}$ remains relatively underexplored. 
A rare example is the as-cast V$_{1-x}$Zr$_{x}$ alloy superconductors reported in 2019\cite{Chandra:JAP2019}.
The highest $J_\mathrm{c}$ was observed in the  $x=$0.4 sample, which consists of $\alpha$-Zr, $\gamma$-ZrV$_{2}$ and $\gamma$'-ZrV$_{2}$. 
The latter two intermetallic compounds belong to the C15 Laves phase family, with slightly differing chemical compositions. 
The eutectic microstructure arises from $\alpha$-Zr and $\gamma$-ZrV$_{2}$. 
The V$_{0.6}$Zr$_{0.4}$ alloy exhibits superconductivity with $T_\mathrm{c}$=8.5 K, $\mu_{0}H_\mathrm{c2}$(0)=17.5 T, and a self-field $J_\mathrm{c}$ of 5$\times$10$^{5}$ A/cm$^{2}$ at 2 K.

In eutectic HEAs, the ability to modify the eutectic microstructure morphology is anticipated to alter superconducting properties and enhance $J_\mathrm{c}$. 
A Japanese research team comprising members from the Fukuoka Institute of Technology, Tokyo Metropolitan University, and Kyushu Sangyo University investigated NbScTiZr, a eutectic HEA known to consist of (NbTi)-enriched bcc and (ScZr)-enriched hcp phases\cite{Rogal:MSEA2016,Krnel:Materials2022}. 
The superconducting properties of as-cast NbScTiZr were studied by Krnel et al.\cite{Krnel:Materials2022}, who reported bulk superconductivity in the bcc phase with $T_\mathrm{c}$=7.3 K. 
The Japanese team further explored the effects of annealing on the fundamental superconducting properties and $J_\mathrm{c}$\cite{Seki:JSNM2023,Kitagawa:MTC2024}.

\begin{figure}
\begin{center}
\includegraphics[width=1\linewidth]{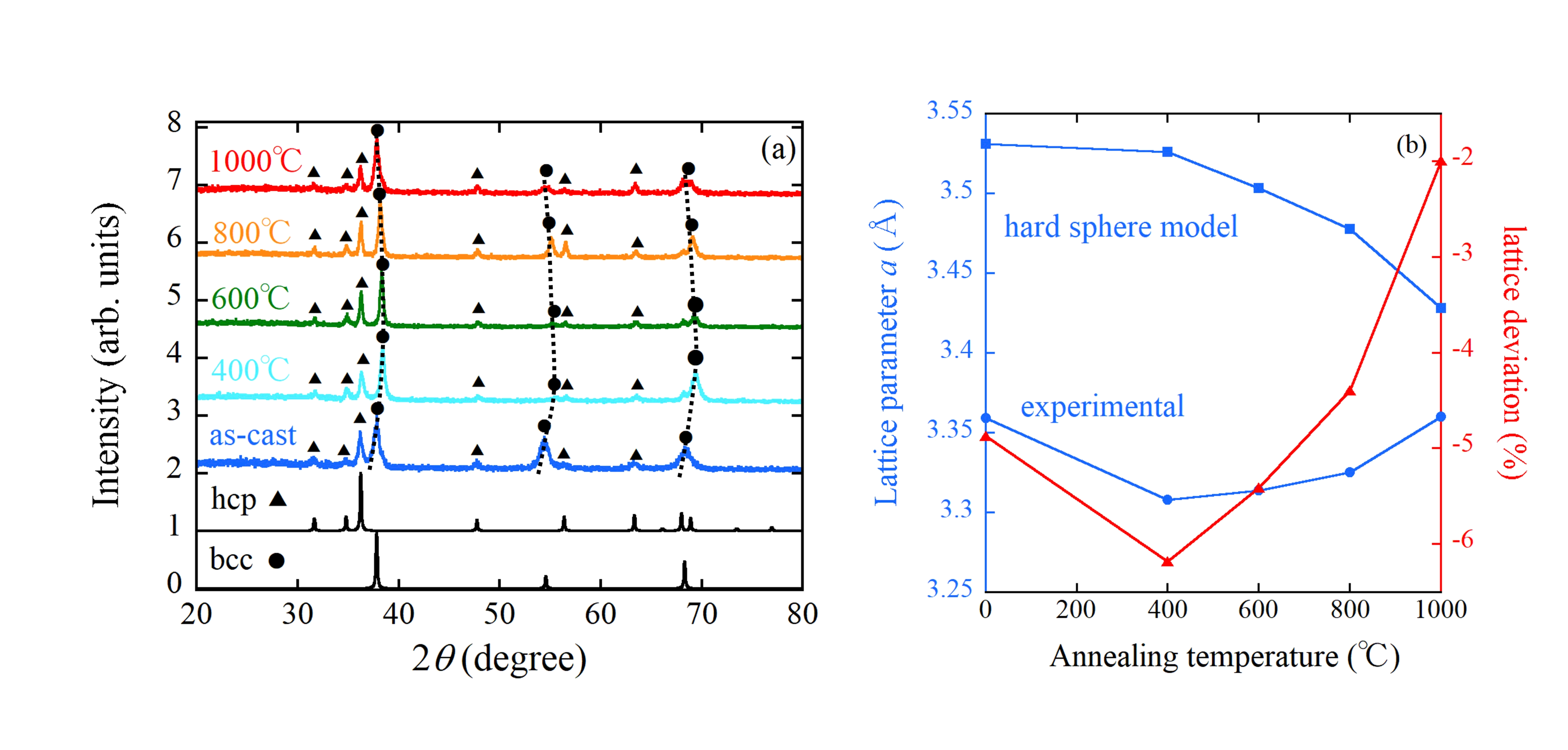}
\caption{\label{fig17}(a) XRD patterns of NbScTiZr samples prepared under different annealing conditions. (b) Dependence of lattice parameter (filled blue circles) and that calculated using the hard-sphere model (filled blue squares) on annealing temperature. The annealing temperature dependence of lattice deviation is also shown. Reproduced with permission from \cite{Kitagawa:MTC2024}.}
\end{center}
\end{figure}

\begin{figure}
\begin{center}
\includegraphics[width=0.8\linewidth]{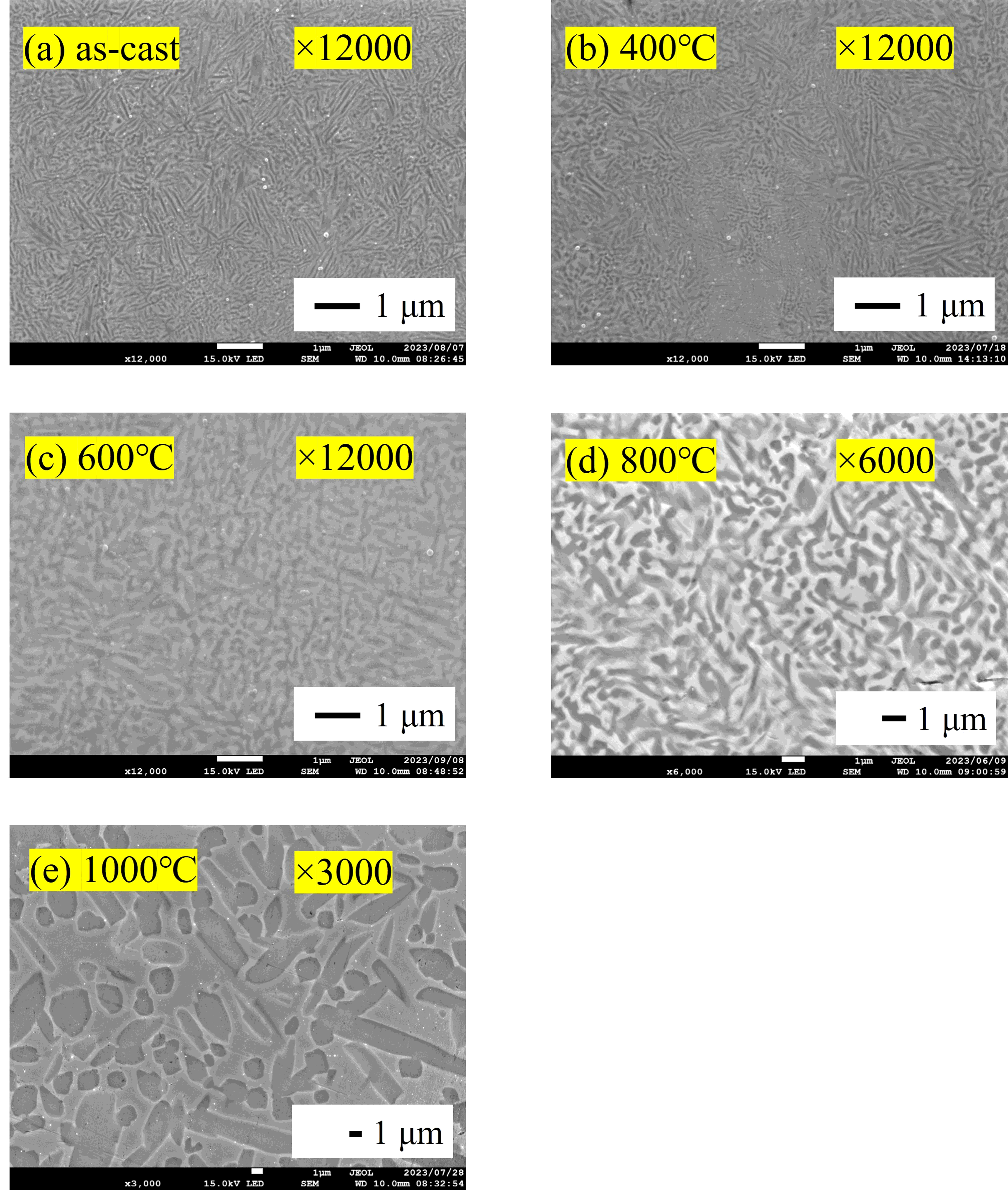}
\caption{\label{fig18} SEM images of NbScTiZr for as-cast sample (a) and heat-treated samples at (b) 400 $^{\circ}$C, (c) 600 $^{\circ}$C, (d) 800 $^{\circ}$C, and (e) 1000 $^{\circ}$C, respectively. Reproduced with permission from \cite{Seki:JSNM2023}.}
\end{center}
\end{figure}

\subsection{Fundamental superconducting properties}
Figure\hspace{1mm}\ref{fig17}(a) depicts the XRD patterns of all prepared samples, which can be interpreted as a superposition of patterns corresponding to the bcc and hcp phases. 
The lattice parameters ($a_\mathrm{exp}$) of the bcc phase are plotted as a function of annealing temperature in Fig.\hspace{1mm}\ref{fig17}(b), with the as-cast sample considered equivalent to annealing at 0 $^{\circ}$C. 
The ideal lattice parameter ($a_\mathrm{hard}$) is calculated using the hard-sphere model, which assumes the absence of lattice strain. 
The lattice strain is then quantified using the lattice deviation, defined as $(a_\mathrm{exp}-a_\mathrm{hard})/a_\mathrm{hard}\times 100$. 
The deviation exhibits a shallow minimum of -6 \% at 400 $^{\circ}$C. 
Notably, the magnitude of this deviation is larger compared to other HEA superconductors, such as HfMoNbTiZr (-1.7 \%)\cite{Kitagawa:JALCOM2022} and (Ti$_{35}$Hf$_{25}$)(Nb$_{25}$Ta$_{5}$)Re$_{10}$ (-2.3 \%)\cite{Hattori:JAMS2023}. 
This suggests that, similar to Ta$_{1/6}$Nb$_{2/6}$Hf$_{1/6}$Zr$_{1/6}$Ti$_{1/6}$ and (TaNb)$_{0.7}$(HfZrTi)$_{0.5}$, thermal annealing induces lattice strain, which appears to be a common characteristic of Nb-based HEAs.
While the origin of the lattice strain induced by thermal annealing in Ta$_{1/6}$Nb$_{2/6}$Hf$_{1/6}$Zr$_{1/6}$Ti$_{1/6}$ and (TaNb)$_{0.7}$(HfZrTi)$_{0.5}$ remains unclear, interface strengthening is responsible for the lattice strain in NbScTiZr\cite{Kitagawa:MTC2024}. 
The interface between the bcc and hcp phases serves as a barrier to dislocation motion due to the mismatch in structural and mechanical properties at the interface. 
The Vickers microhardness, which attains its maximum value at 400 $^{\circ}$C, supports the enhancement of interface strengthening\cite{Kitagawa:MTC2024}.
Figures\hspace{1mm}\ref{fig18}(a)-(e) present scanning electron microscopy (SEM) images showing the eutectic microstructures, where the bright and dark phases correspond to the bcc and hcp phases, respectively. 
In all samples, the volume fraction of the bcc phase is approximately 50 \%. 
The as-cast sample exhibits a lamellar-like microstructure with an average thickness of $\sim$70 nm (Fig.\hspace{1mm}\ref{fig18}(a)), which is retained in the sample annealed at 400 $^{\circ}$C. 
With further increases in annealing temperature, the grain size grows systematically. 
Chemical analysis of the as-cast sample reveals that the bcc phase is Sc-deficient, while the hcp phase is Sc-rich. 
Significant shifts in chemical composition occur in samples annealed above 800 $^{\circ}$C, with the bcc phase exhibiting increased Nb and Ti concentrations.

\begin{figure}
\begin{center}
\includegraphics[width=1\linewidth]{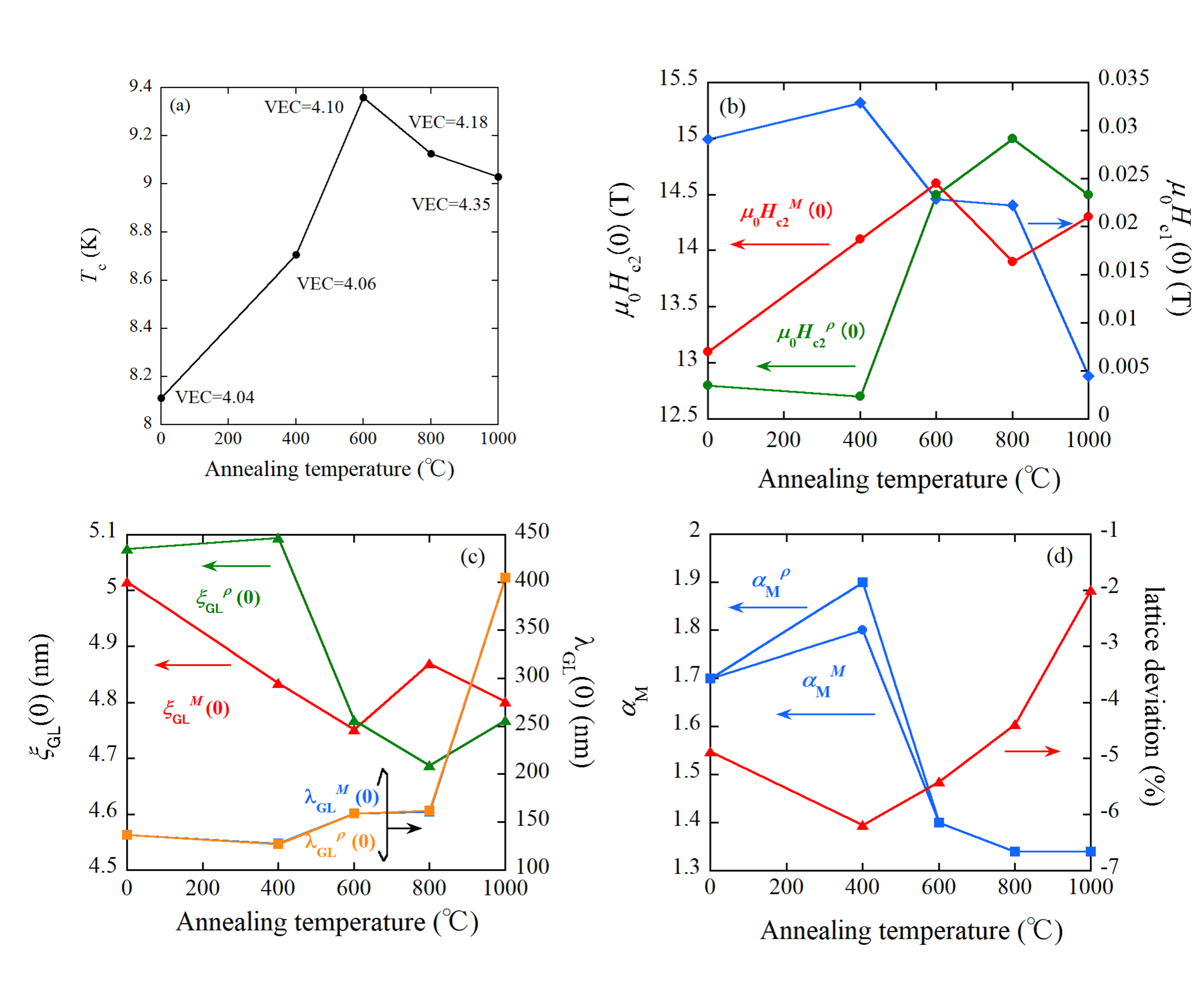}
\caption{\label{fig19} Annealing temperature dependence of (a) $T_\mathrm{c}$, (b) lower critical field $\mu_{0}H_\mathrm{c1}$(0) and $\mu_{0}H_\mathrm{c2}$(0), (c) $\xi_\mathrm{GL}$(0) and $\lambda_\mathrm{GL}$(0), and (d) $\alpha_\mathrm{M}$ for NbScTiZr, respectively. Reproduced with permission from \cite{Kitagawa:MTC2024}.}
\end{center}
\end{figure}

Figure\hspace{1mm}\ref{fig19}(a) shows the dependence of $T_\mathrm{c}$ on annealing temperature. 
The as-cast sample undergoes a superconducting transition at $T_\mathrm{c}$=8.11 K, which increases significantly to 9.36 K in the sample annealed at 600 $^{\circ}$C before slightly decreasing at higher annealing temperatures. 
Figures\hspace{1mm}\ref{fig19}(b) and (c) summarize the annealing temperature dependence of fundamental superconducting parameters. 
The values of $\mu_{0}H_\mathrm{c2}$(0), the magnetic penetration depth ($\lambda_\mathrm{GL}(0)$), and $\xi_\mathrm{GL}(0)$ show no systematic dependence on the annealing temperature. 
The temperature-dependent upper critical field of each sample is fitted using the Werthamer–Helfand–Hohenberg (WHH) model. 
In this fitting, the Maki parameters ($\alpha_\mathrm{M}$) can be derived and are plotted as a function of annealing temperature in Fig.\hspace{1mm}\ref{fig19}(d). 
The $\alpha_\mathrm{M}$ values reach a maximum at 400 $^{\circ}$C, coinciding with the strongest lattice deviation. 
Moreover, the annealing temperature dependence of $\alpha_\mathrm{M}$ aligns well with that of the lattice deviation.
The Maki parameter $\alpha_\mathrm{M}$ is expressed by $\frac{\sqrt{2}\mu_{0}H_\mathrm{c2}^\mathrm{orb}}{\mu_{0}H_\mathrm{c2}^\mathrm{Pauli}}$, where $\mu_{0}H_\mathrm{c2}^\mathrm{Pauli}$ is the Pauli limiting field and $\mu_{0}H_\mathrm{c2}^\mathrm{orb}$ is the orbital-limited upper critical field.
The calculated $\mu_{0}H_\mathrm{c2}^\mathrm{orb}$ values, obtained from magnetization measurements, are 17.9, 20.4, 17.0, 15.4, and 15.3 T, corresponding to the as-cast, 400 $^{\circ}$C, 600 $^{\circ}$C, 800 $^{\circ}$C, and 1000 $^{\circ}$C samples, respectively. 
Similarly, resistivity measurements yield values of 17.9, 21.5, 17.0, 15.4, and 15.3 T for the same samples. These results indicate that $\mu_{0}H_\mathrm{c2}^\mathrm{orb}$ is enhanced at lower annealing temperatures, with larger lattice strain correlating with elevated $\mu_{0}H_\mathrm{c2}^\mathrm{orb}$.
The pronounced lattice strain is associated with the fine eutectic microstructure, which involves significant structural mismatches at the bcc–hcp phase interfaces. 
This combination of lattice strain and structural mismatch strengthens flux pinning, consistent with the results of the $J_\mathrm{c}$ values discussed below. 
Strong flux pinning mitigates Cooper pair destruction caused by Lorentz forces, thereby enhancing $\mu_{0}H_\mathrm{c2}^\mathrm{orb}$.

\begin{figure}
\begin{center}
\includegraphics[width=1\linewidth]{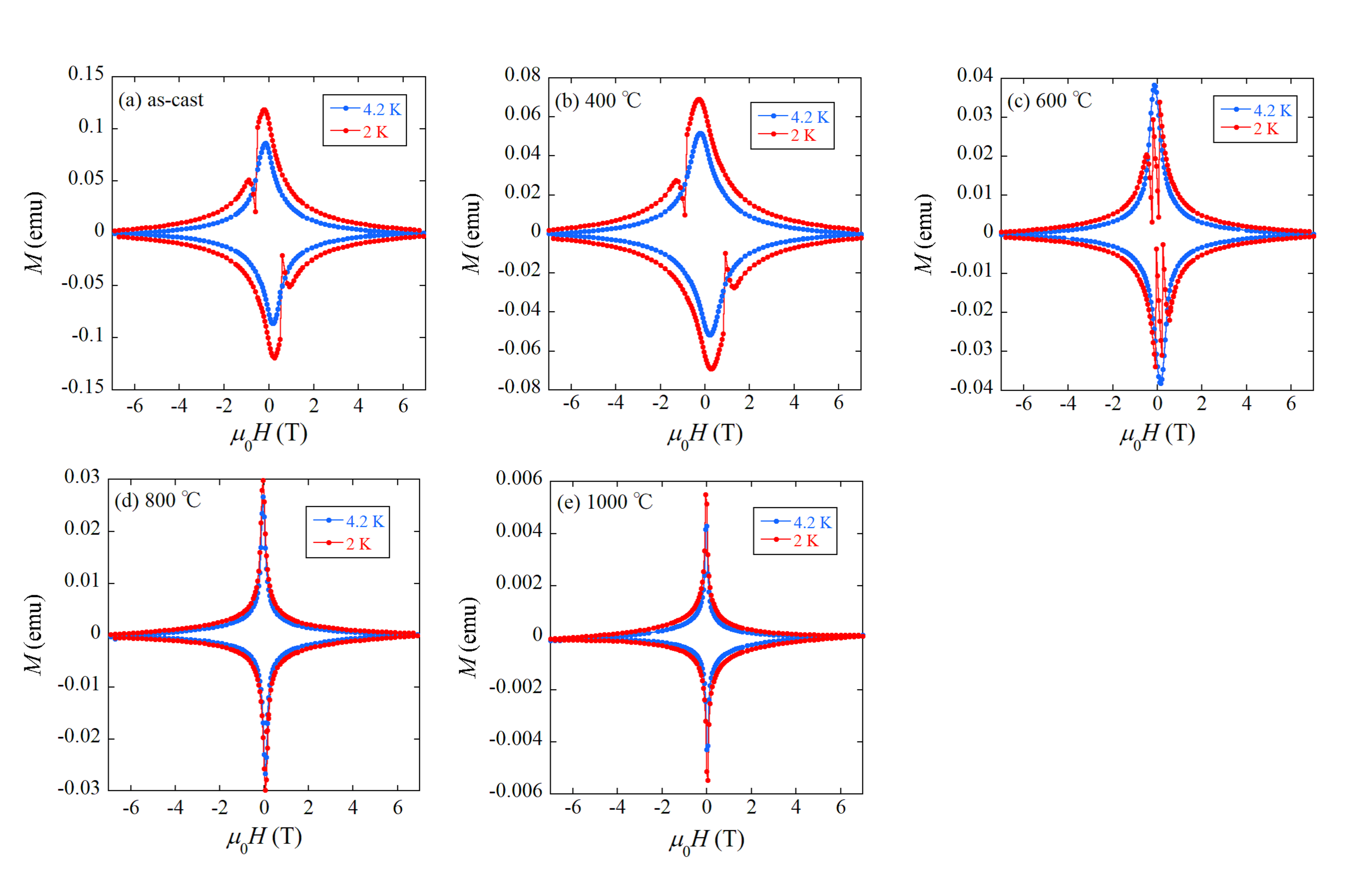}
\caption{\label{fig20} (a), (b), (c), (d), and (e) Isothermal magnetization curves at 2 and 4.2 K for 
NbScTiZr, corresponding to as-cast, 400 $^{\circ}$C annealed, 600 $^{\circ}$C annealed, 800 $^{\circ}$C annealed, and 1000 $^{\circ}$C annealed samples, respectively. Partially reproduced with permission from \cite{Seki:JSNM2023}.}
\end{center}
\end{figure}

\subsection{Annealing effect of $J_\mathrm{c}$}
\begin{figure}
\begin{center}
\includegraphics[width=0.85\linewidth]{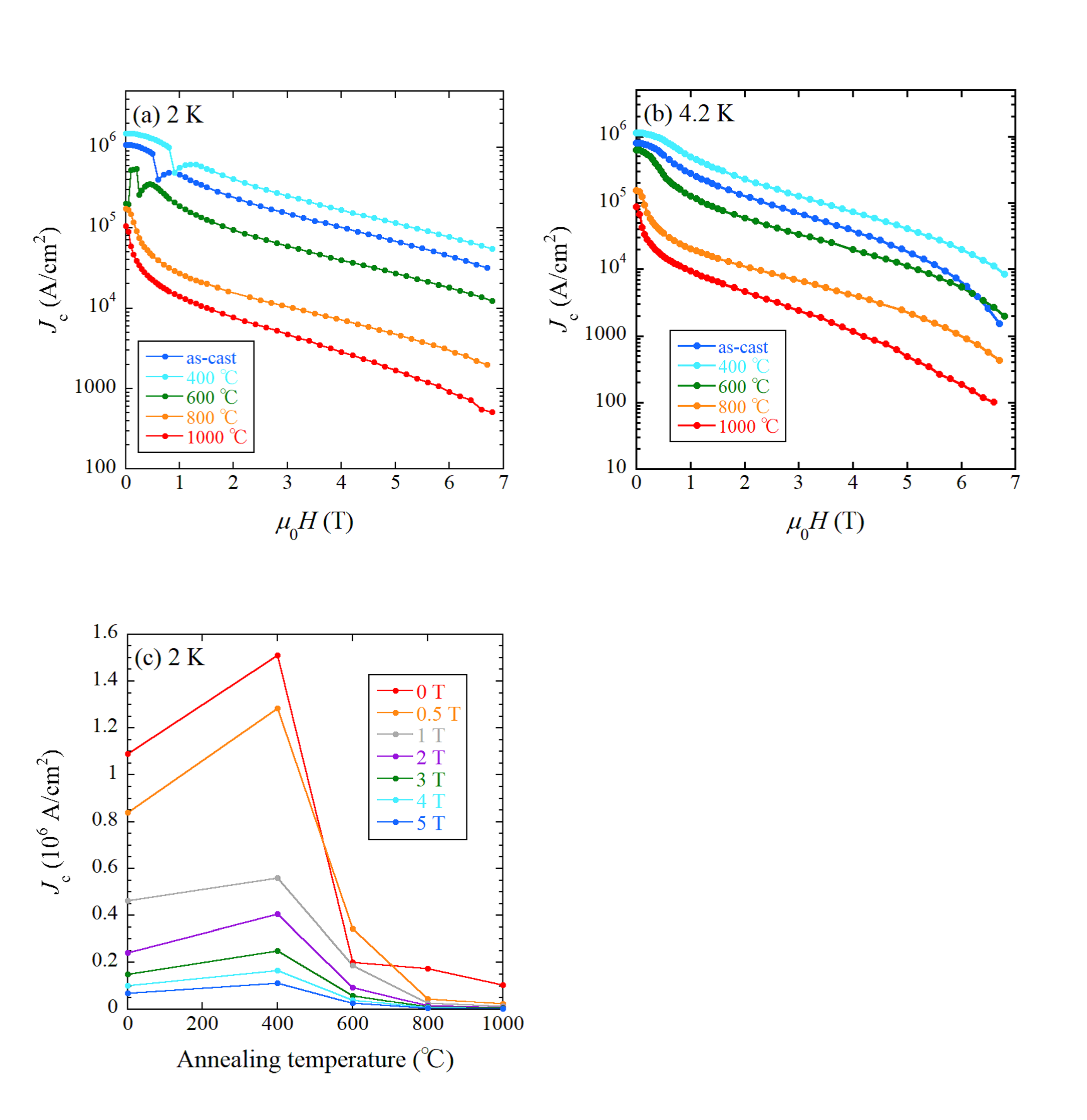}
\caption{\label{fig21} (a) and (b) Magnetic field-dependent $J_\mathrm{c}$ of NbScTiZr at 2 and 4.2 K, respectively. (c) Annealing temperature dependences of $J_\mathrm{c}$ at 2 K under several magnetic fields for NbScTiZr.}
\end{center}
\end{figure}

The results of the isothermal magnetic hysteresis loops at 2 K and 4.2 K are summarized in Figs.\hspace{1mm}\ref{fig20}(a)-(e) for all samples. 
The hysteresis loops of samples annealed at higher temperatures exhibit smaller areas compared to their as-cast and 400 $^{\circ}$C annealed counterparts. 
For the as-cast, 400 $^{\circ}$C, and 600 $^{\circ}$C annealed samples at 2 K, abrupt changes in $M$ occur at low fields, as shown in Fig.\hspace{1mm}\ref{fig20}(a), (b), and (c), respectively. 
In contrast, no such abrupt variations in $M$ are discernible at 4.2 K in the same figures. 
This behavior is attributed to the flux jump phenomenon commonly observed in various superconductors, including the aforementioned HEAs\cite{Jung:NC2022,Gao:APL2022}.
Figures\hspace{1mm}\ref{fig21}(a) and (b) illustrate the magnetic field dependence of $J_\mathrm{c}$ at 2 K and 4.2 K, respectively. 
Notably, the self-field $J_\mathrm{c}$ at 2 K for the as-cast and 400 $^{\circ}$C annealed samples exceeds 1 MA/cm$^{2}$, classifying these materials as possessing exceptionally high $J_\mathrm{c}$ values. 
Under a fixed magnetic field, the $J_\mathrm{c}$ of the as-cast sample slightly increases with annealing up to 400 $^{\circ}$C, but rapidly diminishes by an order of magnitude with further increases in annealing temperature (see Fig.\hspace{1mm}\ref{fig21}(c)). 
This trend strongly correlates with the lattice strain, which reaches its maximum at 400 $^{\circ}$C, as shown in Fig.\ref{fig17}(b). 
Consequently, the larger lattice strain observed at lower annealing temperatures induces higher $J_\mathrm{c}$, similar to the behavior reported for heat-treated Ta$_{1/6}$Nb$_{2/6}$Hf$_{1/6}$Zr$_{1/6}$Ti$_{1/6}$ and (TaNb)$_{0.7}$(HfZrTi)$_{0.5}$.
A significant additional finding in NbScTiZr is the relationship between $J_\mathrm{c}$ and the eutectic microstructure. 
As observed in Figs.\hspace{1mm}\ref{fig18}(a) and (b), the eutectic microstructure of the as-cast or 400 $^{\circ}$C annealed sample is notably fine. 
However, annealing temperatures exceeding 400 $^{\circ}$C result in a systematic coarsening of the morphology. 
Grain boundaries, which act as flux pinning centers, are more abundant in the fine eutectic microstructure. 
This increased density of grain boundaries between the bcc and hcp phases contributes to the higher $J_\mathrm{c}$ observed in the as-cast and 400 $^{\circ}$C annealed samples.
In summary, the enhancement of $J_\mathrm{c}$ in NbScTiZr is attributed to two key factors: the lattice strain induced by thermal annealing and the fine eutectic microstructure. 

\begin{figure}
\begin{center}
\includegraphics[width=1\linewidth]{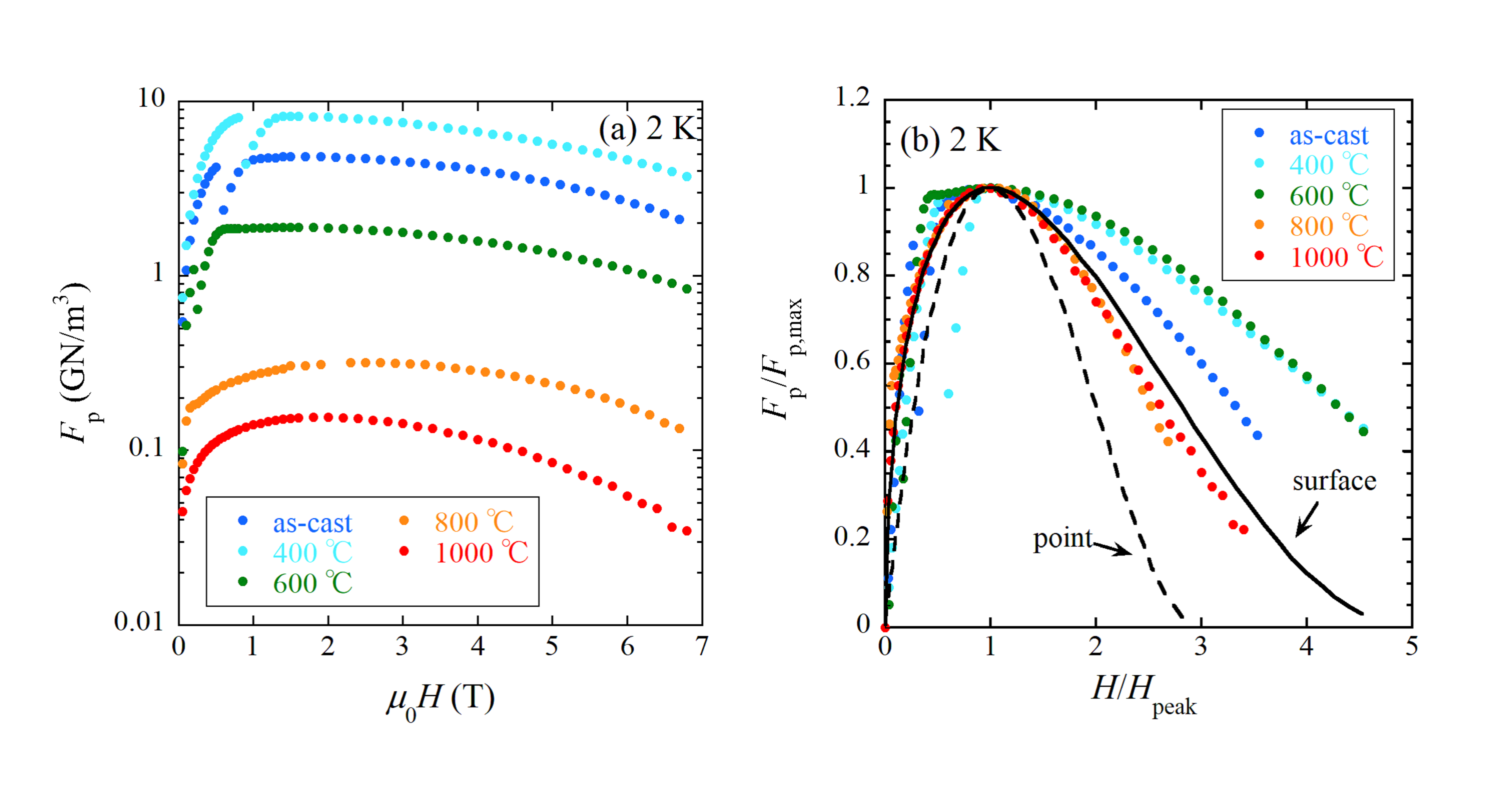}
\caption{\label{fig22} (a) Field-dependent flux pinning force densities, and (b) normalized flux pinning force densities as a function of $H/H_\mathrm{peak}$ for NbScTiZr samples at 2 K.}
\end{center}
\end{figure}

To elucidate the flux pinning mechanism, the flux pinning force densities ($F_\mathrm{p}$) for all samples at 2 K are calculated and plotted as a function of the external field in Fig.\hspace{1mm}\ref{fig22}(a). 
Notably, $F_\mathrm{p}$ reaches nearly 10 GN/m$^{3}$ at approximately 1 T in the 400 $^{\circ}$C annealed sample. 
In both the as-cast and 400 $^{\circ}$C annealed samples, the maximum $F_\mathrm{p}$ occurs at approximately 1 T, whereas for the 800 $^{\circ}$C and 1000 $^{\circ}$C annealed samples, this maximum shifts to higher magnetic fields. 
At the intermediate annealing temperature of 600 $^{\circ}$C, a shoulder-like feature appears around 0.5 T, suggesting a possible crossover in the pinning mechanism between the low- and high-temperature annealed samples.
Figure\hspace{1mm}\ref{fig22}(b) presents an analysis of the pinning mechanism using the normalized flux pinning force density ($F_\mathrm{p}/F_\mathrm{p,max}$) as a function of $H/H_\mathrm{peak}$. 
The magnetic field dependencies of $F_\mathrm{p}/F_\mathrm{p,max}$ for samples subjected to high-temperature annealing (800 $^{\circ}$C and 1000 $^{\circ}$C) are well described by the surface pinning model. 
In these samples, the eutectic structure size is relatively large, and the lattice strain is weak. 
The surface pinning model suggests that the interface between the bcc and hcp phases predominantly serves as the pinning site. 
In contrast, the samples annealed at lower temperatures exhibit substantial deviations from both the surface and point pinning models, implying that an alternative mechanism or a combination of mechanisms may be required to fully account for the flux pinning behavior, which is influenced by lattice strain and the fine eutectic microstructure.

\section{(Nb,Zr,Pt)$_{67}$Ti$_{33}$}\label{sec6}
\begin{figure}
\begin{center}
\includegraphics[width=0.6\linewidth]{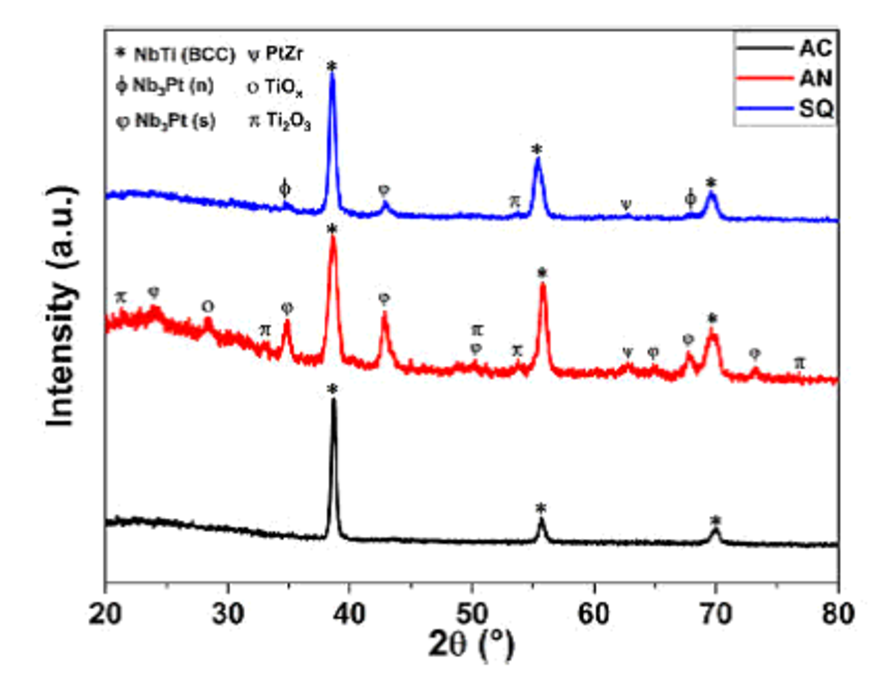}
\caption{\label{fig23} XRD patterns of (Nb,Zr,Pt)$_{67}$Ti$_{33}$ samples prepared under various annealing conditions. AC, AN, and SQ denote the as-cast, 1100 $^{\circ}$C annealed, and solutionized + quenched samples, respectively. Reproduced with permission from \cite{Srivastava:JMCC2024}.}
\end{center}
\end{figure}

\begin{figure}
\begin{center}
\includegraphics[width=1\linewidth]{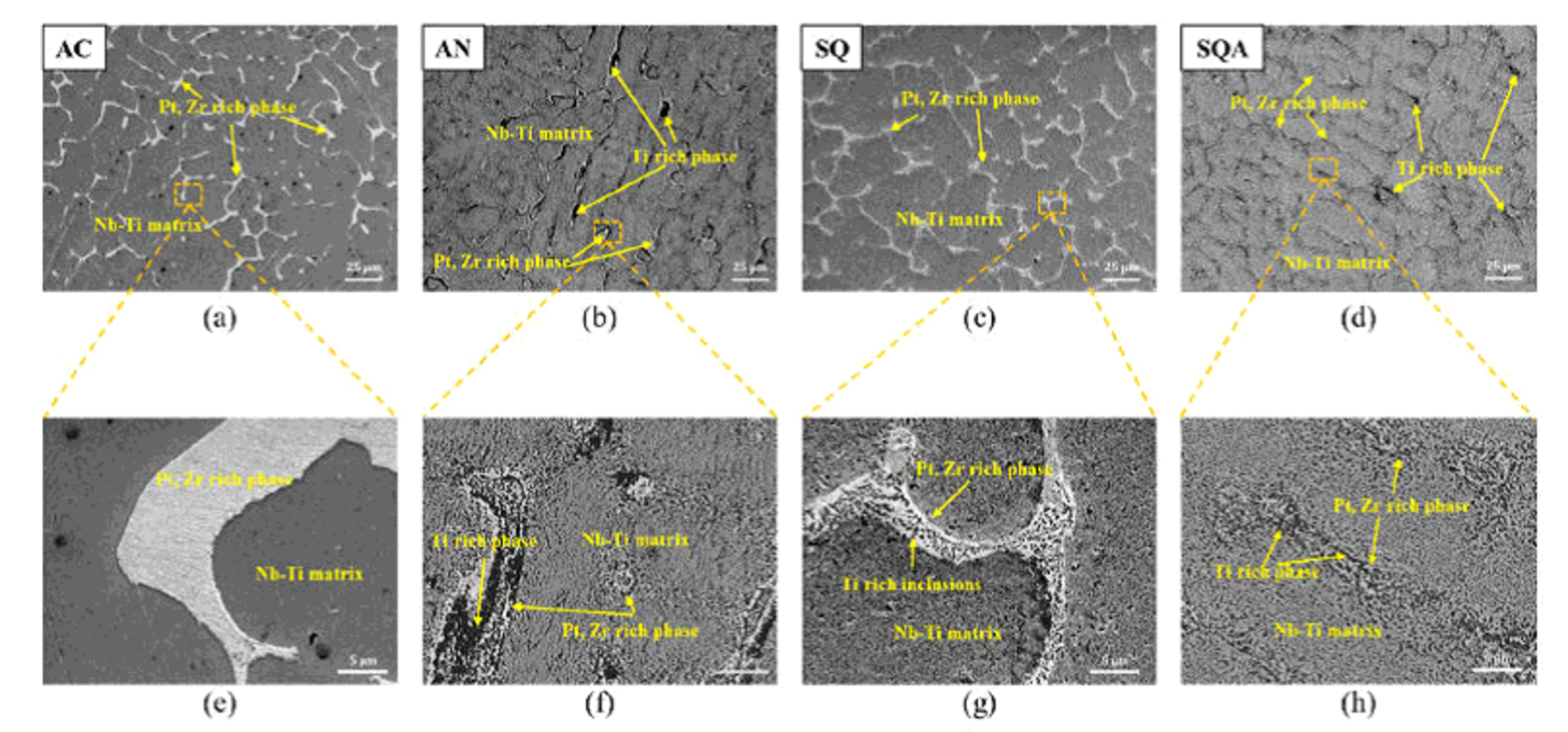}
\caption{\label{fig24} SEM images of (Nb,Zr,Pt)$_{67}$Ti$_{33}$ samples: (a) AC (as-cast), (b) AN (annealing at 1100 $^{\circ}$C for 9h), (c) SQ (solutionizing for 9h + quenched), and (d) SQA (solutionized for 9h + quenched + annealing at 1100 $^{\circ}$C for 9h), respectively. (e)–(h) Enlarged views of (a)-(d). Reproduced with permission from \cite{Srivastava:JMCC2024}.}
\end{center}
\end{figure}

Srivastava et al. investigated the correlation between $J_\mathrm{c}$ and microstructure in the multiphase (Nb,Zr,Pt)$_{67}$Ti$_{33}$ HEA superconductor\cite{Srivastava:JMCC2024}.
Although this alloy is multiphase, each impurity phase does not include $\alpha$-Ti, which is present in Nb-Ti alloys. 
Therefore, the effect of impurity species on $J_\mathrm{c}$ performance can be studied over a broader range.
The composition of (Nb, 5 at.\% Zr, 5 at.\% Pt)$_{67}$Ti$_{33}$ was synthesized using the arc-melting method. 
Samples with varying annealing conditions, denoted as AC, AN, SQ, and SQA, were prepared. 
The AC sample received no heat treatment (as-cast). 
The heat treatment protocols for the other samples were as follows: AN—annealing at 1100 $^{\circ}$C for 9 hours; SQ—solutionizing for 9 hours followed by quenching; and SQA—solutionizing for 9 hours, quenching, and subsequent annealing at 1100 $^{\circ}$C for 9 hours.
Figure\hspace{1mm}\ref{fig23} presents the XRD patterns of the AC, AN, and SQ samples. 
In all samples, the primary phase is the bcc structure, identified as the Nb-Ti phase through chemical composition analysis. 
Additionally, AN and SQ samples exhibit stoichiometric (denoted as s) and non-stoichiometric (n) phases of A15 Nb$_{3}$Pt. 
SEM images (Figs.\hspace{1mm}\ref{fig24}(a)-(h)) reveal eutectic regions in all samples except AC. 
Elemental mapping indicates that Pt and Zr are uniformly distributed throughout the matrix in the AN, SQ, and SQA samples.

\begin{figure}
\begin{center}
\includegraphics[width=1\linewidth]{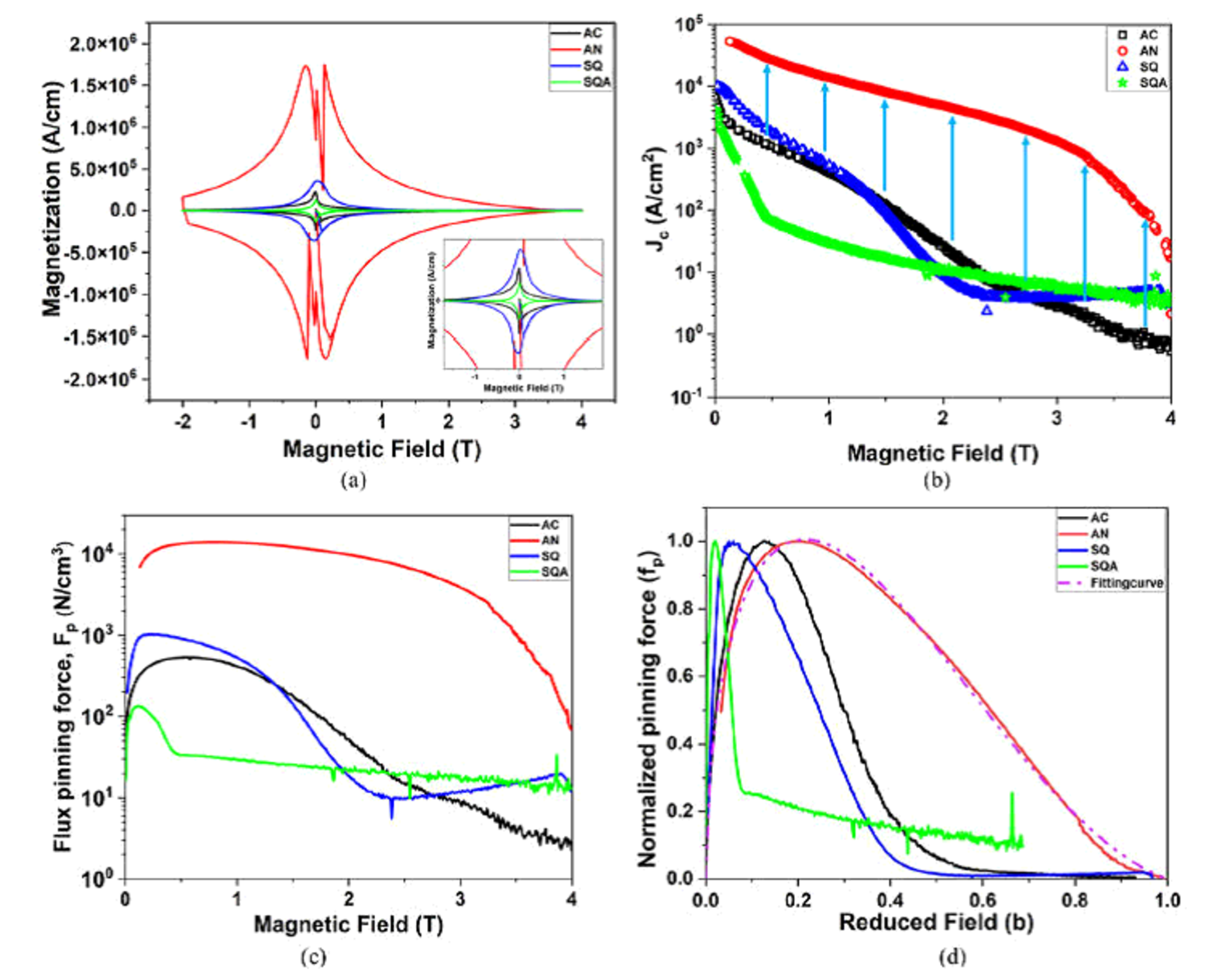}
\caption{\label{fig25} (a) Magnetic hysteresis loops at 4.2 K, (b) magnetic field dependences of $J_\mathrm{c}$, (c) flux pinning force densities, and (d) normalized flux pinning force densities of (Nb,Zr,Pt)$_{67}$Ti$_{33}$ samples. Reproduced with permission from \cite{Srivastava:JMCC2024}.}
\end{center}
\end{figure}

The critical temperatures of the AC, AN, SQ, and SQA samples are 8.1 K, 8.3 K, 7.8 K, and 5.6 K, respectively. 
A two-step superconducting transition is observed in the AC and SQ samples, indicating some chemical inhomogeneity. 
The authors suggest that this phenomenon is related to the formation of the non-stoichiometric Nb$_{3}$Pt phase in the AC and SQ samples. 
In the SQA sample, which exhibits the lowest $T_\mathrm{c}$, the growth of the s phase at elevated annealing temperatures is inferred.
The $M$–$H$ hysteresis loops measured at 4.2 K are shown in Fig.\hspace{1mm}\ref{fig25}(a). 
The AN sample exhibits a significantly larger hysteresis loop area compared to the other samples. 
Figure\hspace{1mm}\ref{fig25}(b) compares the magnetic field performance of $J_\mathrm{c}$ for all samples, with the AN sample demonstrating the best performance, achieving a self-field $J_\mathrm{c}$ of 80 kA/cm$^{2}$ at 4.2 K. 
The higher $J_\mathrm{c}$ in the AN sample is attributed to a greater fraction of the stoichiometric Nb$_{3}$Pt phase alongside the Nb-Ti phase.
Another possible contributor to the high $J_\mathrm{c}$ is the magnetic flux pinning provided by impurity phases (e.g., Ti$_{2}$Zr, $\beta$-NbZr, Pt-Zr) within the matrix.
The magnetic flux pinning mechanism is analyzed in Figs.\hspace{1mm}\ref{fig25}(c) and (d). 
The flux pinning force of the AN sample is one to two orders of magnitude higher compared to the other samples. 
The functional form of the magnetic field dependence of the flux pinning force is shown to be dependent on the annealing conditions. 
This suggests that the pinning mechanism may vary from sample to sample.
The authors employed the Dew-Hughes model, expressed as $f_\mathrm{p}(h)=A_{1}h^{x_{1}}(1-h)^{y_{1}}+A_{2}h^{x_{2}}(1-h)^{y_{2}}$. 
The fitting parameters obtained for the AN sample are $x_{1}$ = 0.5, $y_{1}$ =2, $x_{2}$ = 0.5 and $y_{2}$ = 1.
The first term corresponds to the surface normal pinning model, while the second term represents the magnetic normal pinning model\cite{Dew-Hughes:PhyMag1974}.

\section{Comparison of $J_\mathrm{c}$ among different HEAs including Nb-Ti alloy}\label{sec7}
\begin{figure}
\begin{center}
\includegraphics[width=1\linewidth]{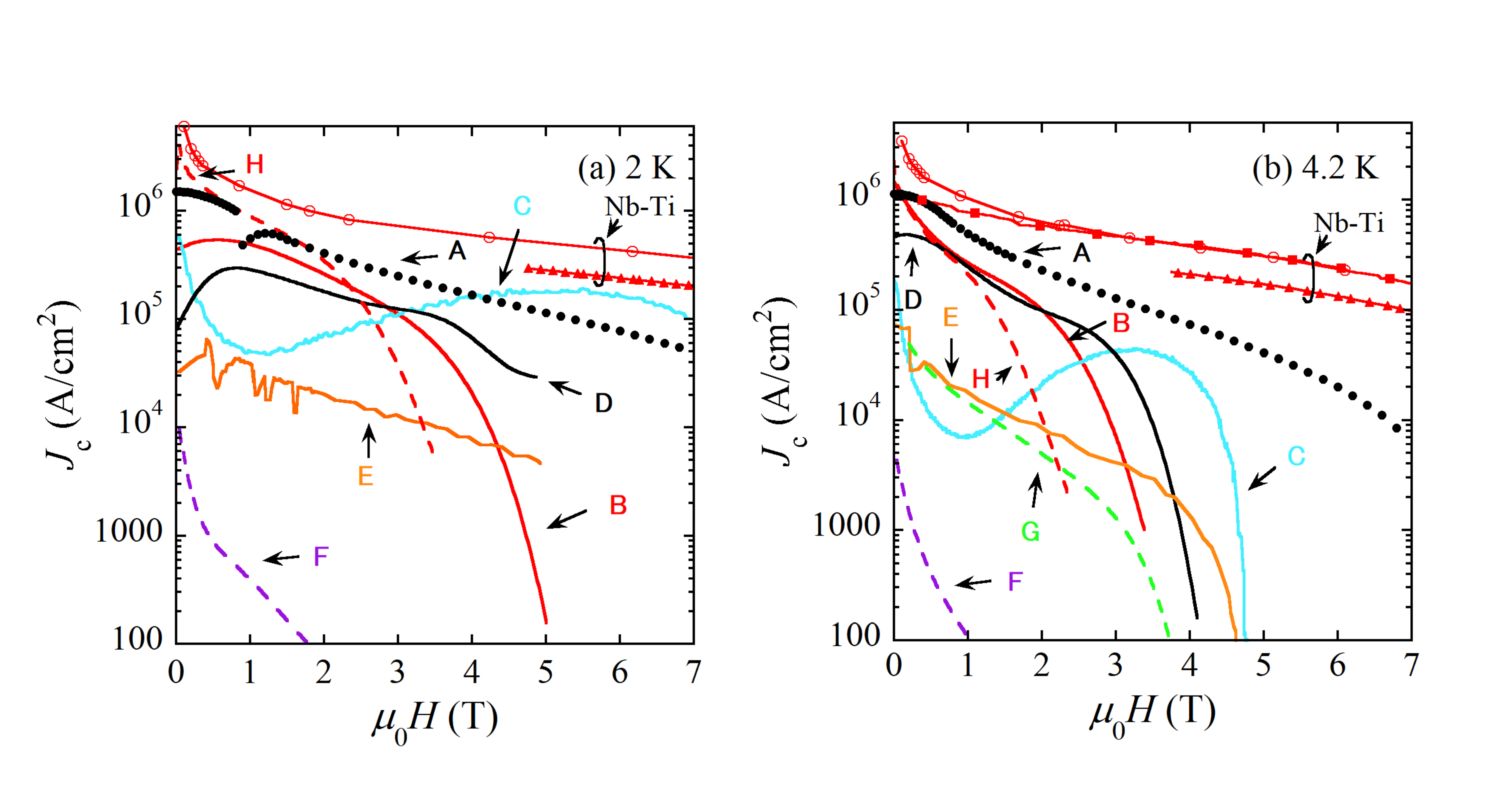}
\caption{\label{fig26} Comparison of $J_\mathrm{c}$ among different HEAs including Nb-Ti alloys (a) at 2 K and (b) at 4.2 K, respectively. A: NbScTiZr annealed at 400 $^{\circ}$C, B: Ta$_{1/6}$Nb$_{2/6}$Hf$_{1/6}$Zr$_{1/6}$Ti$_{1/6}$ thin film ($T_\mathrm{s}$=520 $^{\circ}$C) fabricated using a target prepared by a ball-milling followed by hot-press sintering\cite{Jung:NC2022}, C: (TaNb)$_{0.7}$(HfZrTi)$_{0.5}$ annealed at 500 $^{\circ}$C\cite{Gao:APL2022}, D: Ta$_{1/6}$Nb$_{2/6}$Hf$_{1/6}$Zr$_{1/6}$Ti$_{1/6}$ annealed at 550 $^{\circ}$C\cite{Kim:JMST2024}, E: Ta$_{1/6}$Nb$_{2/6}$Hf$_{1/6}$Zr$_{1/6}$Ti$_{1/6}$ SPS\cite{Kim:AM2022}, F: Ta$_{1/6}$Nb$_{2/6}$Hf$_{1/6}$Zr$_{1/6}$Ti$_{1/6}$ as-cast\cite{Kim:AM2020}, G: (Nb,Zr,Pt)$_{67}$Ti$_{33}$ annealed at 1100 $^{\circ}$C\cite{Srivastava:JMCC2024}, H: Ta$_{1/6}$Nb$_{2/6}$Hf$_{1/6}$Zr$_{1/6}$Ti$_{1/6}$ thin film fabricated using targets after thermal annealing at 700 $^{\circ}$C\cite{Jung:PSAC2024}. Red triangles: commercial Nb–47 wt.\%Ti alloy reported in the 1980s\cite{Collings:book}, Red squares: Nb-Ti multifilamentary strands reported in 1985\cite{Larbalestier:IEEE1996}, Red open circles: state-of-the-art Nb-Ti wire used in accelerators\cite{Boutboul:IEEE2006}.}
\end{center}
\end{figure}

Table\hspace{1mm}\ref{tab1} summarizes the fundamental superconducting properties, microstructural features, and the presence of lattice strain in representative HEAs discussed in this review. 
Figures\hspace{1mm}\ref{fig26}(a) and (b) exhibit the comparison of $J_\mathrm{c}$ among selected HEAs at 2 K and 4.2 K, respectively. 
The thin-film form of Ta$_{1/6}$Nb$_{2/6}$Hf$_{1/6}$Zr$_{1/6}$Ti$_{1/6}$ generally enhances $J_\mathrm{c}$ at low fields\cite{Jung:NC2022,Jung:PSAC2024}. 
If the high-field performance of $J_\mathrm{c}$ can be substantially improved, thin-film HEAs hold promise for practical applications.
Hereafter, the focus shifts to bulk HEAs. 
The overall magnetic field performance of $J_\mathrm{c}$ in as-cast or SPS-processed Ta$_{1/6}$Nb$_{2/6}$Hf$_{1/6}$Zr$_{1/6}$Ti$_{1/6}$\cite{Kim:AM2022,Kim:AM2020} is relatively inferior compared to heat-treated Ta$_{1/6}$Nb$_{2/6}$Hf$_{1/6}$Zr$_{1/6}$Ti$_{1/6}$, (TaNb)$_{0.7}$(HfZrTi)$_{0.5}$, and NbScTiZr\cite{Gao:APL2022,Seki:JSNM2023,Kim:JMST2024}. 
As shown in Table\hspace{1mm}\ref{tab1}, the as-cast and SPS-processed Ta$_{1/6}$Nb$_{2/6}$Hf$_{1/6}$Zr$_{1/6}$Ti$_{1/6}$ lack lattice strain. 
This comparison underscores the critical role of lattice strain in enhancing $J_\mathrm{c}$.
Although both NbScTiZr and (Nb,Zr,Pt)$_{67}$Ti$_{33}$ exhibit a eutectic microstructure, the former demonstrates superior $J_\mathrm{c}$ performance. 
In (Nb,Zr,Pt)$_{67}$Ti$_{33}$, lattice strain is not detected\cite{Srivastava:JMCC2024}, whereas significant strain is observed in NbScTiZr. 
Thus, comparing these two HEAs further highlights the pivotal role of lattice strain. 
While the low-field $J_\mathrm{c}$ (1 $<$ $\mu_{0}H$ $<$ 2 T) of (TaNb)$_{0.7}$(HfZrTi)$_{0.5}$ at 2 K is an order of magnitude lower than that of heat-treated Ta$_{1/6}$Nb$_{2/6}$Hf$_{1/6}$Zr$_{1/6}$Ti$_{1/6}$, it exhibits relatively higher $J_\mathrm{c}$ values at higher fields ($\mu_{0}H$ $>$ 3 T), attributed to the fishtail effect.
At 4.2 K, the $J_\mathrm{c}$ of NbScTiZr annealed at 400 $^{\circ}$C surpasses all other bulk HEAs within the measured field range. 
At 2 K, the $J_\mathrm{c}$ of 400 $^{\circ}$C-annealed NbScTiZr attains the highest value between 0 and 4 T among bulk HEAs but declines in comparison to (TaNb)$_{0.7}$(HfZrTi)$_{0.5}$ at fields above 4 T. 
These comparative evaluations highlight the potential of eutectic HEA superconductors to achieve high $J_\mathrm{c}$ values.
Lattice strain is detected in (TaNb)$_{0.7}$(HfZrTi)$_{0.5}$ and NbScTiZr, contributing to the enhancement of $J_\mathrm{c}$ at lower fields. 
The fine eutectic microstructure in NbScTiZr, featuring an abundance of grain boundaries, further enhances low-field $J_\mathrm{c}$, making it the highest-performing bulk HEA in this regard. 
In the (TaNb)$_{0.7}$(HfZrTi)$_{0.5}$ system, the fishtail effect arises from the precipitation of nano-sized phases ($\sim$10 nm), which significantly improves $J_\mathrm{c}$ at higher fields.
Conversely, the absence of nano-sized precipitates in Ta$_{1/6}$Nb$_{2/6}$Hf$_{1/6}$Zr$_{1/6}$Ti$_{1/6}$ and NbScTiZr leads to rapid $J_\mathrm{c}$ degradation under elevated fields. 
Therefore, the synergy between eutectic microstructure with lattice strain and nano-sized precipitates offers a promising pathway to achieving superior $J_\mathrm{c}$ values across a wide range of magnetic fields.

\begin{table}[h]
\caption{\label{tab1}Comparison of $T_\mathrm{c}$, $\mu_{0}H_\mathrm{c2}$(0), structural features among Ta$_{1/6}$Nb$_{2/6}$Hf$_{1/6}$Zr$_{1/6}$Ti$_{1/6}$ samples prepared by various methods, (TaNb)$_{0.7}$(HfZrTi)$_{0.5}$, NbScTiZr, and (Nb,Zr,Pt)$_{67}$Ti$_{33}$.}%
\begin{tabular}{@{}lllll@{}}
\toprule
HEA    & $T_\mathrm{c}$ (K) & $\mu_{0}H_\mathrm{c2}$(0) (T)  & microstructure & lattice strain  \\
\midrule
Ta$_{1/6}$Nb$_{2/6}$Hf$_{1/6}$Zr$_{1/6}$Ti$_{1/6}$ & & & & \\
\hspace{2mm}as-cast & 7.85 & 12.05 & single phase & no \\
\hspace{2mm}SPS & 7.83 & 10.5 & impurity phases & no \\
    &      &      & (Hf$_{2}$Fe, ZrFe$_{2}$) & \\
\hspace{2mm}thin film & 7.28 & $\sim$12 & single phase & no \\
\hspace{2mm}($T_\mathrm{s}$=520 $^{\circ}$C) & & & & \\
\hspace{2mm}annealed at 550 $^{\circ}$C & $\sim$7.8 & $\sim$11.5 & single phase & yes \\
(TaNb)$_{0.7}$(HfZrTi)$_{0.5}$ & & & & \\
\hspace{2mm}annealed at 500 $^{\circ}$C & $\sim$7 & 10.07 & clusters $\sim$10 nm & yes \\
NbScTiZr & & & & \\
\hspace{2mm}annealed at 400 $^{\circ}$C & 8.71 & 12.7 or 14.1 & eutectic microstructure & yes \\
(Nb,Zr,Pt)$_{67}$Ti$_{33}$ & & & & \\
\hspace{2mm}annealed at 1100 $^{\circ}$C & 8.3 & - & eutectic microstructure & no \\
\botrule
\end{tabular}
\end{table}

In considering the application as a superconducting wire, a comparison of $T_\mathrm{c}$, $\mu_{0}H_\mathrm{c2}$(0), and $J_\mathrm{c}$ between HEAs and Nb-Ti alloys is essential. 
The $T_\mathrm{c}$ and $\mu_{0}H_\mathrm{c2}$(0) values of the NbTi alloy are 10 K and 14.5 T, respectively\cite{Slimani:book}. 
Although the $\mu_{0}H_\mathrm{c2}$(0) value of NbScTiZr is comparable to that of the NbTi alloy, slightly lower $\mu_{0}H_\mathrm{c2}$(0) values are reported for many HEAs (see Table\hspace{1mm}\ref{tab1}). 
The $T_\mathrm{c}$ value of NbScTiZr is the highest among the HEAs; however, it remains 1 to 2 K lower than that of the Nb-Ti alloy. 
Hence, the pursuit of enhancing $T_\mathrm{c}$ without compromising $\mu_{0}H_\mathrm{c2}$(0) emerges as the next area of research. 
Figures\hspace{1mm}\ref{fig26}(a) and (b) also depict the magnetic field dependence of $J_\mathrm{c}$ for Nb-Ti superconductors.
The red triangles denote $J_\mathrm{c}$ data for a commercial Nb–47 wt.\%Ti alloy reported in the 1980s\cite{Collings:book}.
The red squares and open circles represent the field-dependent $J_\mathrm{c}$ of Nb-Ti multifilamentary strands\cite{Larbalestier:IEEE1996} and the state-of-the-art Nb-Ti wire employed in accelerators\cite{Boutboul:IEEE2006}, respectively. 
At 2 K, NbScTiZr and (TaNb)$_{0.7}$(HfZrTi)$_{0.5}$ exhibit $J_\mathrm{c}$ values comparable to those of the 1980s commercial Nb–47 wt.\%Ti alloy within the presented magnetic field range. 
The low-field $J_\mathrm{c}$ values at 4.2 K for NbScTiZr are nearly identical to those of Nb-Ti multifilamentary strands. 
However, a pronounced deviation occurs as the magnetic field increases, likely attributable to the lower $T_\mathrm{c}$ of NbScTiZr relative to the Nb-Ti alloy and differences in sample morphology (bulk: NbScTiZr, multifilamentary strands: Nb-Ti). 
To achieve $J_\mathrm{c}$ values comparable to those of the state-of-the-art Nb-Ti wire used in accelerators, an enhancement of at least an order of magnitude is required across both low and high-field regimes. 
Therefore, further research efforts, including in-depth investigations of wire fabrication methodologies, are necessary in addition to studies that optimize flux pinning strength.

\section{Future perspectives}\label{sec8}
The lattice strain, manifested as a reduction in lattice parameters, and the eutectic microstructure are effective strategies for enhancing $J_\mathrm{c}$ at lower magnetic fields. 
Conversely, nanoscale precipitates serve as flux-pinning sites under higher magnetic fields. 
However, direct observation of magnetic flux distribution under both low and high fields using techniques such as magneto-optical microscopy and scanning tunneling microscopy is essential.

In Ta$_{1/6}$Nb$_{2/6}$Hf$_{1/6}$Zr$_{1/6}$Ti$_{1/6}$ and (TaNb)$_{0.7}$(HfZrTi)$_{0.5}$, lattice strain is induced by thermal annealing, contrasting with NbScTiZr, which exhibits lattice strain inherently due to its eutectic microstructure. 
The as-cast NbScTiZr sample already possesses considerable lattice strain, as evidenced by deviations from the hard-sphere model. 
This strain likely originates from the structural mismatch between the bcc and hcp phases.
While the origin of lattice strain in Ta$_{1/6}$Nb$_{2/6}$Hf$_{1/6}$Zr$_{1/6}$Ti$_{1/6}$ and (TaNb)$_{0.7}$(HfZrTi)$_{0.5}$ remains unclear, possible microstructural anomalies warrant detailed metallographic investigation. 
Furthermore, determining whether lattice strain exists in other superconducting HEAs is interesting, necessitating further exploration of annealing effects in HEAs with elemental combinations beyond those discussed in this review.

In eutectic HEA superconductors, the quinary systems remain underexplored. 
It is anticipated that introducing an additional element into NbScTiZr could reduce $\xi_\mathrm{GL}(0)$, thereby elevating $\mu_{0}H_\mathrm{c2}$(0), which is advantageous for practical applications as superconducting wires. 
A finer eutectic microstructure generally correlates with a higher critical current. 
Achieving thermally induced eutectic microstructures in single-phase HEA superconductors in the as-cast state could potentially yield ultra-fine microstructures through precise annealing temperature control.

For improving $J_\mathrm{c}$ at higher magnetic fields, the nanoscale precipitates observed in (TaNb)$_{0.7}$(HfZrTi)$_{0.5}$ and Nb-Ti alloys are promising. 
However, designing such precipitates poses significant challenges. 
In Ta$_{1/6}$Nb$_{2/6}$Hf$_{1/6}$Zr$_{1/6}$Ti$_{1/6}$ and Nb$_{2/5}$Hf$_{1/5}$Zr$_{1/5}$Ti$_{1/5}$ samples synthesized via ball-milling and SPS methods, high-field $J_\mathrm{c}$ performance improves relative to arc-melted samples, attributed to Fe-based intermetallic precipitates serving as flux-pinning sites. 
The Fe source is the stainless-steel jar and balls used in ball milling. 
Similar artificial pinning sites have been reported in Nb-Ti alloys. 
For instance, Mousavi et al. investigated $J_\mathrm{c}$ in Nb-Ti synthesized by high-energy ball milling of Nb, Ti, and Y$_{2}$O$_{3}$ powders, where heat treatment at 800 $^{\circ}$C precipitated Y$_{2}$Ti$_{2}$O$_{7}$ particles (2–5 nm), achieving $J_\mathrm{c}$$\sim$5$\times$10$^{5}$ A/cm$^{2}$ at 5 T\cite{Mousavi:MD2021}. 
Thus, the ball-milling method holds the potential for intentionally introducing nanoscale pinning sites in HEAs.

\begin{figure}
\begin{center}
\includegraphics[width=0.6\linewidth]{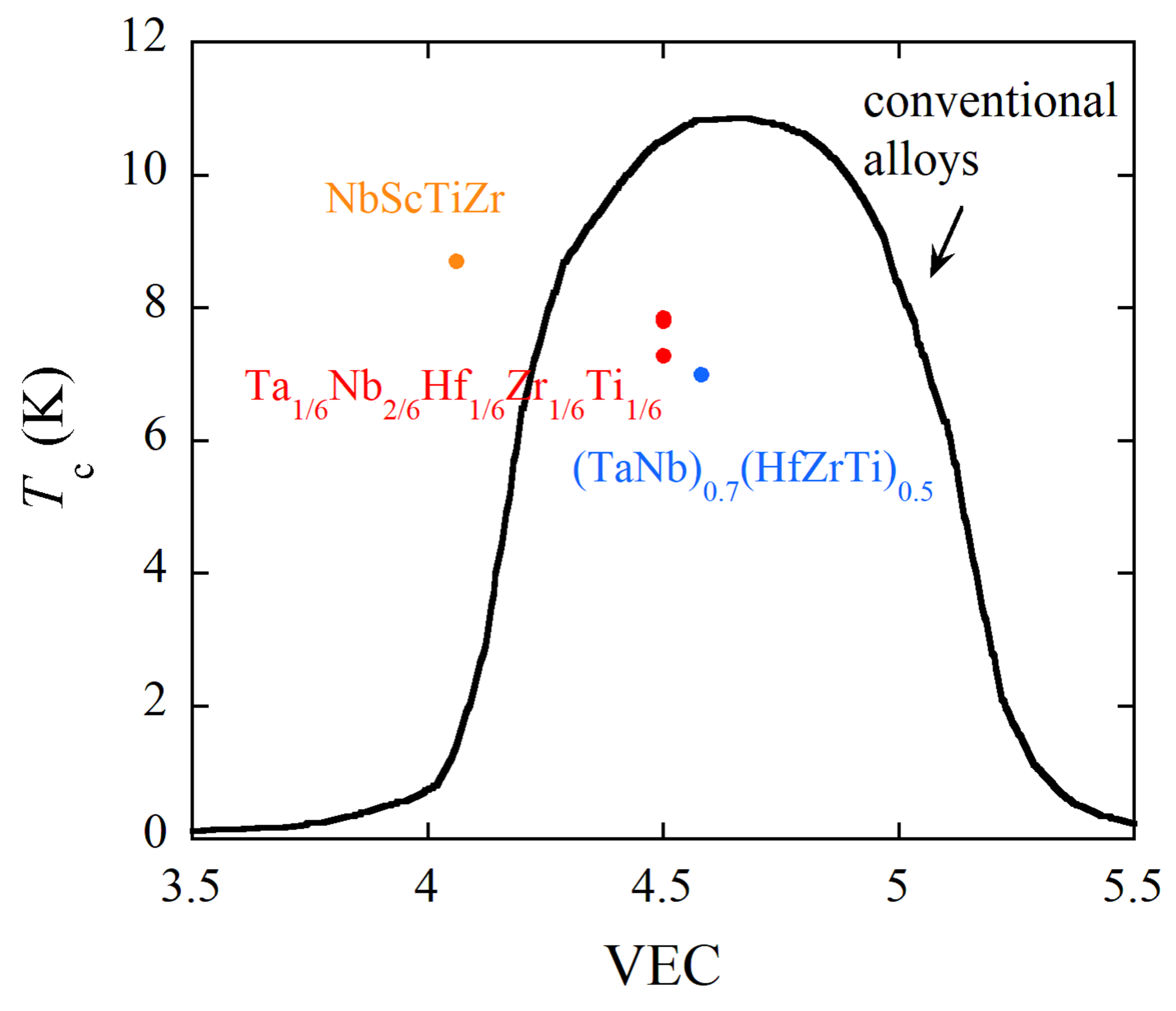}
\caption{\label{fig27} Variation of $T_\mathrm{c}$ with VEC for
Ta$_{1/6}$Nb$_{2/6}$Hf$_{1/6}$Zr$_{1/6}$Ti$_{1/6}$, (TaNb)$_{0.7}$(HfZrTi)$_{0.5}$, and NbScTiZr listed in Table\hspace{1mm}\ref{tab1}. The solid line represents the Matthias rule for conventional binary and ternary transition metal-based bcc alloys.}
\end{center}
\end{figure}

Higher $T_\mathrm{c}$ values than those reported in this review are essential for practical superconducting applications.
Empirically, bcc superconducting alloys follow the Matthias rule, suggesting that $T_\mathrm{c}$ strongly correlates with the density of states, represented by valence electron concentration (VEC)\cite{Matthias:PR1955}. 
Conventional binary and ternary bcc alloys exhibit a universal trend described by the solid line in Fig.\hspace{1mm}\ref{fig27}. 
Results for selected HEAs from this review are also plotted. Notably, the actual bcc phase composition of NbScTiZr annealed at 400 $^{\circ}$C is Nb$_{27.9}$Sc$_{21.6}$Ti$_{25.3}$Zr$_{25.2}$\cite{Seki:JSNM2023}.
While quinary HEA superconductors display Matthias-rule-like behavior, their $T_\mathrm{c}$ values are reduced compared to conventional alloys, as observed for Ta$_{1/6}$Nb$_{2/6}$Hf$_{1/6}$Zr$_{1/6}$Ti$_{1/6}$ and (TaNb)$_{0.7}$(HfZrTi)$_{0.5}$. 
However, an enhancement in $T_\mathrm{c}$ is evident in NbScTiZr with its fine eutectic microstructure, underscoring the importance of materials research on eutectic HEA superconductors with varied VECs.

Only one research paper reports the feasibility of HEA superconducting wire\cite{Jung:JALCOM2024}, as discussed in subsection \ref{wire}. 
At this stage, the $J_\mathrm{c}$ values of wire samples remain unsatisfactory, particularly with regard to their magnetic performance at higher fields. 
To address this, optimizing flux pinning—such as by introducing nanoscale precipitates—proves to be a promising approach. 
Another critical factor influencing $J_\mathrm{c}$ is the morphology of the grain boundary. 
The reported method for fabricating HEA superconducting wires is currently limited to sintering. 
For instance, drawing a rod of HEA produced via arc melting could significantly alter the morphology of the grain boundary.

For practical deployment, machinability and cost-effectiveness of constituent elements are crucial. 
The mechanical properties of HEAs discussed in this review still need to be explored, necessitating future investigations. 
Since Hf and Sc are relatively expensive compared to Ti, Zr, and Nb, reducing their content should be a priority in the design of superconducting wires.

\section{Summary}\label{sec9}
This review surveyed studies on $J_\mathrm{c}$ in various bcc HEA superconductors. 
The Korean team has extensively researched Ta$_{1/6}$Nb$_{2/6}$Hf$_{1/6}$Zr$_{1/6}$Ti$_{1/6}$ in bulk, thin-film, and wire forms. 
The SPS-prepared samples exhibit relatively higher $J_\mathrm{c}$ compared to their as-cast counterparts, primarily attributed to Fe-based intermetallic impurities. 
The Fe contamination originates from the stainless-steel jar and balls utilized in the ball-milling process. 
Remarkable improvements in $J_\mathrm{c}$ are achieved in thin-film and thermally treated samples.
While the thin-film state generally enhances $J_\mathrm{c}$, the lattice strain induced by thermal annealing is pivotal in strengthening flux pinning in heat-treated samples.
The (TaNb)$_{0.7}$(HfZrTi)$_{0.5}$ system is particularly intriguing due to its pronounced fishtail effect, which increases $J_\mathrm{c}$ with elevated magnetic fields. 
This effect is attributed to nanoscale precipitates. 
Additionally, lattice strain, observed at optimal annealing temperatures, contributes to improving the low-field $J_\mathrm{c}$ performance. 
The eutectic NbScTiZr alloy demonstrates enhanced $J_\mathrm{c}$, driven by the combined effects of lattice strain and its fine eutectic microstructure. 
In contrast to Ta$_{1/6}$Nb$_{2/6}$Hf$_{1/6}$Zr$_{1/6}$Ti$_{1/6}$ and (TaNb)$_{0.7}$(HfZrTi)$_{0.5}$, lattice strain in NbScTiZr is inherent in the as-cast state. 
However, thermal annealing at 400 $^{\circ}$C introduces even greater strain. 
This suggests that lattice strain induced by thermal annealing may be a common feature of Nb-based HEAs.

We also compared $J_\mathrm{c}$ among representative HEAs discussed in this review. 
Employing both eutectic microstructures accompanied by lattice strain and nanosized precipitates will likely achieve significantly elevated $J_\mathrm{c}$ values over a broad magnetic field range.
Finally, we addressed future perspectives: understanding the origin of lattice strain, exploring finer eutectic microstructures, artificially introducing nanoscale pinning sites, and enhancing $T_\mathrm{c}$. 
For practical applications as superconducting wires, prioritizing improvements in high-field $J_\mathrm{c}$ performance and investigating the mechanical properties of HEAs are critical tasks.

\bmhead{Acknowledgments}

J.K. acknowledges the support from a Grant-in-Aid for Scientific Research (KAKENHI) (Grant No. 23K04570). Y.M. acknowledges the support from a Grant-in-Aid for Scientific Research (KAKENHI) (Grant No. 21H00151). T.N. acknowledges the support from a Grant-in-Aid for Scientific Research (KAKENHI) (Grant Nos. 20K03867 and 24K08236) and the Takahashi Industrial and Economic Research Foundation. 

\section*{Author contributions}
All the authors contributed equally to the study and preparation of the paper.

\section*{Declarations}

\bmhead{Conflict of interest}
The authors declare that they have no conflict of interest.

\end{document}